\documentclass[a4paper,11pt]{article}
\pdfoutput=1 % if your are submitting a pdflatex (i.e. if you have
\usepackage{jcappub} % for details on the use of the package, please
\usepackage[T1]{fontenc} % if needed

\usepackage{graphicx}
\title{Gravitational waves from inclined orbits around Kerr black holes}

\renewcommand{\thefootnote}{\fnsymbol{footnote}}
\author[a]{Yi-Han Huang}
\author[a,*]{Sen Guo}
\author[b]{Yu Liang}
\author[a,*]{Lin Wen}
\author[c]{Kai Lin}
\affiliation[a]{College of Physics and Electronic Engineering, Chongqing Normal University, Chongqing 401331, People's Republic of China}
\affiliation[b]{School of Big Data and Artificial Intelligence, Fuyang University of Technology, Fuyang 236000, People's Republic of China}
\affiliation[c]{Universidade Federal de Campina Grande, Campina Grande, PB, Brasil, \\ Instituto de F\'isica, Universidade de S\~ao Paulo, S\~ao Paulo, Brasil}

\renewcommand{\thefootnote}{\fnsymbol{footnote}}
\footnotetext[1]{Corresponding author.}
\renewcommand{\thefootnote}{\arabic{footnote}}
\emailAdd{sguophys@126.com, wlqx@cqnu.edu.cn}

\abstract{Extreme mass ratio inspirals on generic inclined orbits provide a critical dynamical probe of strong field gravity, as they uniquely break the severe parameter degeneracies inherent to planar equatorial trajectories. To isolate the exact rotational signatures, we extend the periodic-orbit framework to three-dimensional inclined trajectories in the Kerr spacetime, specifically focusing on the $r$--$\theta$ orbital resonances. By mapping the geodesic motion into a co-precessing spatial framework parameterized by Mino time, we compute the emitted quadrupole waveforms and evolve the local dynamics using circularity-preserving radiation reaction. It is demonstrated that frame-dragging induces a transient dynamical tilt at periastron, which directly splits the emitted radiation into an exact five-comb spectral architecture. When this resonant structure is resolved over a single coherence time, the statistical error on the dimensionless spin is reduced to $\sigma_a \approx 10^{-2}$, improving upon the equatorial baseline by up to two orders of magnitude. Because any bound orbit in a static, spherically symmetric spacetime generates a maximum of three frequency combs, the detection of this five-comb hierarchy establishes a rigorous, model-independent null test to strictly exclude spherical geometries.}

\begin{document}
\maketitle
\flushbottom
\renewcommand{\thefootnote}{\arabic{footnote}}

\section{Introduction}
\label{sec:intro}
\par
Space-borne interferometers, such as LISA, Taiji, and TianQin, will observe extreme mass ratio inspirals (EMRIs) composed of a stellar mass compact object orbiting a supermassive black hole~\cite{LISA:2017pwj,Hu:2017mde,TianQin:2015yph}. Over $10^{4}$--$10^{5}$ orbital cycles in the strong field regime, the emitted gravitational waveform maps the background spacetime geometry, enabling measurements of the central mass and spin to fractional precisions of $10^{-6}$--$10^{-4}$~\cite{Babak:2017tow,Amaro-Seoane:2012lgq}, provided that accurate waveform models~\cite{Barack:2018yvs,Chua:2020stf} and robust structural observables are available~\cite{Barausse:2020rsu}. Since astrophysical black holes are expected to rotate~\cite{Reynolds:2020jwt}, their exterior geometry is uniquely described by the Kerr metric~\cite{Carter:1971zc,Robinson:1975bv}. While electromagnetic probes of this geometry, such as black hole shadows and lensing, depend on specific accretion flow and emission models~\cite{Jiang:2023img,Kumar:2025ueq}, gravitational waves provide a direct dynamical probe. EMRIs are expected to form via many-body relaxation in dense nuclear clusters and consequently arrive with essentially random orbital inclinations~\cite{Amaro-Seoane:2007osp}. For such generic orbits, frame dragging causes the orbital plane to precess around the black hole spin axis~\cite{Bardeen:1972fi}. The gravitational waves emitted by inclined Kerr orbits therefore offer a direct measurement of strong-field rotation. Periodic orbits, which naturally generate exact frequency combs, provide an ideal analytical framework to isolate this rotational signature.

\par
The commensurability between radial and angular frequencies restricts bound geodesics to a discrete set of periodic orbits, uniquely classified by a topological integer triplet $(z,w,v)$ representing their zoom--whirl structure~\cite{Levin:2008mq,Grossman:2011ps}. The gravitational wave signatures of these orbits have been extensively investigated across various static, spherically symmetric spacetimes, including quantum-corrected, regular, and hairy black holes. In these scenarios, the modified gravity parameters shift the innermost stable and marginally bound circular orbits, thereby altering the specific energy and angular momentum required to excite a given $(z,w,v)$ resonance~\cite{Tu:2023xab,Zhao:2024exh,Junior:2024tmi,Alloqulov:2025bxh,Deng:2025wzz,Alloqulov:2025dqi,He:2026txc,Alloqulov:2026jqm}. This geometric shift directly modifies the timing and amplitude of the zoom--whirl bursts~\cite{Yang:2024lmj,Choudhury:2025qsh,Zahra:2025tdo,Ahmed:2025azu,Gogoi:2026obd}. Although Fisher information analyses can estimate the detectability of such deviations, different static spacetimes frequently yield nearly indistinguishable planar orbits~\cite{Fu:2024cfk,Gong:2025mne,Huang:2026oga}. Furthermore, extensions of this framework to rotating spacetimes have primarily been restricted to the equatorial plane, where the black hole spin and alternative-gravity parameters shift the exact same dynamical quantities~\cite{Li:2025sfe,Yang:2026syx,Jumaniyozov:2026hrt,Shahzad:2026hcc,Dai:2026ujj}. While off-equatorial periodic orbits have been dynamically classified~\cite{Yang:2026syx,Dai:2026ujj}, their waveforms remain largely unexplored. In a recent study~\cite{Huang:2026oga}, we demonstrated a fundamental limitation of equatorial trajectories: the azimuthal frequency shift induced by frame dragging is highly degenerate with the orbital parameters. Consequently, a Schwarzschild template can reproduce an equatorial Kerr periodic signal to a mismatch of $10^{-6}$--$10^{-4}$. Over a single coherence time, the equatorial frequency comb alone cannot reliably measure the black hole spin.

\par
For non-equatorial motion, the Carter constant guarantees the integrability of the geodesic equations~\cite{Carter:1968rr}, and the trajectory acquires a third, polar frequency~\cite{Schmidt:2002qk}. Crucially, frame dragging induces a nodal precession of the orbital plane. The weak-field expansion of this precession frequency begins with the Lense--Thirring dipole term and is followed by the mass quadrupole term, allowing orbits at distinct inclinations to break the degeneracy between the two~\cite{Bardeen:1972fi,Ryan:1995wh}. Such orbital precession is a universal astrophysical observable, detected in the weak field via the LAGEOS satellites, Gravity Probe B~\cite{Ciufolini:2004rq,Everitt:2011hp}, and binary pulsars~\cite{VenkatramanKrishnan:2020pbi}. In the Galactic Center, it manifests through the periastron shift of the S2 star~\cite{GRAVITY:2020gka}, where future measurements of orbital planes aim to test the no-hair theorem~\cite{Will:2007pp}. Around accreting compact objects, precession is routinely inferred from quasiperiodic oscillations~\cite{Stella:1997tc,Ingram:2009vm,Liu:2023vfh}, jet dynamics~\cite{Cui:2023uyb}, and the recurrence times of quasiperiodic eruptions driven by star-disk collisions~\cite{Linial:2023nqs,Zhou:2024bjt}. In the context of gravitational waves, the polar and nodal motions introduce harmonics of the form $m\Omega_{\varphi}+j\Omega_{\theta}+n\Omega_{r}$~\cite{Drasco:2005kz}, which modulate the observed strain~\cite{Barack:2003fp}. The geometric distinction is absolute: in every static, spherically symmetric spacetime, the orbital plane is strictly fixed, enforcing $\Omega_{\theta}=|\Omega_{\varphi}|$. The emergence of an independent nodal frequency directly proves the axisymmetry of the background spacetime. Orbits with commensurate radial and polar frequencies---corresponding to the transient resonances crossed during an EMRI evolution~\cite{Flanagan:2010cd,Ruangsri:2013hra}---strictly close within the precessing plane~\cite{Grossman:2011ps}. These $r$--$\theta$ periodic orbits thus serve as the natural theoretical laboratory to isolate the nodal frequency.

\par
In this paper, we compute the gravitational waves emitted by inclined periodic orbits in the Kerr spacetime. We demonstrate that because an $r$--$\theta$ periodic orbit is exactly closed in a co-precessing frame, its quadrupole radiation splits into five distinct frequency combs, separated exactly by multiples of the nodal frequency. To determine whether this spectral signature of rotation survives detector noise, radiation reaction, and full parameter estimation, we confront these waveforms with static Schwarzschild templates. We evolve the orbital parameters via radiation reaction~\cite{Kennefick:1995za,Gair:2005ih}, generate the signal using the numerical kludge framework~\cite{Babak:2006uv}, and perform both a Fisher matrix analysis~\cite{Vallisneri:2007ev} and a direct injection into simulated detector data~\cite{Jaranowski:2009zz}. Furthermore, we derive an analytical lower bound on the waveform mismatch against any planar-orbit template from a static, spherically symmetric metric, evaluated using the distinguishability criterion of Lindblom et al.~\cite{Lindblom:2008cm}. Our results indicate that over a single coherence time at $200$~Mpc, the inclined measurement reduces the statistical error on the spin to $\sigma_{a}\approx 10^{-2}$, an improvement of up to two orders of magnitude over the equatorial baseline. For a dimensionless spin of $a=0.5$, an orbital inclination greater than $8.6^{\circ}$ (in LISA) or $4.4^{\circ}$ (in Taiji) is sufficient to definitively exclude all static, spherical spacetimes. At inclinations above $45^{\circ}$, the nodal frequency can be extracted directly from the data stream without relying on matched filtering templates.

\par
The remainder of this paper is organized as follows. Section~\ref{sec:orbits} derives the inclined Kerr geodesics, the fundamental frequencies, and the nodal precession, establishing the conditions for $r$--$\theta$ periodic orbits. Section~\ref{sec:waves} computes the waveforms and discrete line spectra, evaluates the radiation reaction timescales, and performs the data analysis over one coherence time. Section~\ref{sec:conclusions} summarizes our conclusions.

\section{Geodesic dynamics and inclined periodic orbits in Kerr spacetime}
\label{sec:orbits}
\par
We derive a dynamical observable that vanishes for any static, spherically symmetric black hole but is uniquely determined by the Kerr spin parameter in the strong field regime. The nodal precession of the orbital plane, driven by frame dragging, provides this signature.

\subsection{Mino-time evolution and the co-precessing spatial framework}
\label{sec:geodesics}
\par
In this section, we use dimensionless variables, with lengths and times measured in units of $GM/c^{2}$ and $GM/c^{3}$, respectively, where $M$ is the black-hole mass. Physical units are restored in Sec.~\ref{sec:waves}. In Boyer--Lindquist coordinates, the Kerr metric is given by~\cite{Kerr:1963ud,Boyer:1966qh}
\begin{eqnarray}
\mathrm{d}s^{2} &&= -\Bigl(1-\frac{2r}{\Sigma}\Bigr)\mathrm{d}t^{2} - \frac{4ar\sin^{2}\theta}{\Sigma}\,\mathrm{d}t\,\mathrm{d}\varphi + \frac{\Sigma}{\Delta}\,\mathrm{d}r^{2} + \Sigma\,\mathrm{d}\theta^{2} \nonumber \\
&&+\Bigl(r^{2}+a^{2}+\frac{2a^{2}r\sin^{2}\theta}{\Sigma}\Bigr)\sin^{2}\theta\, \mathrm{d}\varphi^{2},
\label{eq:kerr}
\end{eqnarray}
where $\Sigma=r^{2}+a^{2}\cos^{2}\theta$, $\Delta=r^{2}-2r+a^{2}=(r-r_{+})(r-r_{-})$, and $r_{\pm}=1\pm\sqrt{1-a^{2}}$. Conserved quantities govern the geodesic motion. Alongside the energy $E=-u_{t}$ and axial angular momentum $L_{z}=u_{\varphi}$, the Kerr metric admits the Carter constant~\cite{Carter:1968rr}
\begin{equation}
Q=u_{\theta}^{2}+\cos^{2}\theta\,\Bigl[a^{2}(1-E^{2})+\frac{L_{z}^{2}}{\sin^{2}\theta}\Bigr].
\label{eq:carter}
\end{equation}
In the static limit ($a=0$), $Q=L^{2}-L_{z}^{2}$ represents the squared projection of the orbital angular momentum onto the equatorial plane. For a rotating black hole ($a\neq0$), $Q$ arises from an irreducible Killing tensor, making generic orbits non-planar~\cite{Walker:1970un}. Such orbits probe geometric structures inaccessible to equatorial trajectories~\cite{Gair:2007kr}.

\par
The separability of the Hamilton--Jacobi equation guarantees the integrability of the motion~\cite{Carter:1968rr}
\begin{equation}
\Sigma\,g^{\mu\nu}p_{\mu}p_{\nu} = -\frac{\bigl[(r^{2}+a^{2})\,p_{t}+a\,p_{\varphi}\bigr]^{2}}{\Delta} + \Delta\,p_{r}^{2} + \frac{\bigl(p_{\varphi}+a\sin^{2}\theta\,p_{t}\bigr)^{2}}{\sin^{2}\theta} + p_{\theta}^{2}.
\label{eq:hj}
\end{equation}
Setting $p_{t}=-E$ and $p_{\varphi}=L_{z}$ and introducing the Mino time $\lambda$ ($\mathrm{d}\tau=\Sigma\,\mathrm{d}\lambda$), the equations of motion decouple~\cite{Mino:2003yg}
\begin{eqnarray}
&&\Bigl(\frac{\mathrm{d}r}{\mathrm{d}\lambda}\Bigr)^{2} = R(r),
~~~~~~~~~~~~\Bigl(\frac{\mathrm{d}\zeta}{\mathrm{d}\lambda}\Bigr)^{2} = \Theta_{\zeta}(\zeta), \nonumber \\
&&\frac{\mathrm{d}t}{\mathrm{d}\lambda} = T_{r}(r)+T_{\zeta}(\zeta),
~~~~~~~\frac{\mathrm{d}\varphi}{\mathrm{d}\lambda} = \Phi_{r}(r)+\Phi_{\zeta}(\zeta),
\label{eq:mino}
\end{eqnarray}
where $\zeta\equiv\cos\theta$, $P(r)\equiv E(r^{2}+a^{2})-aL_{z}$, $\beta\equiv a^{2}(1-E^{2})$, and the effective potentials are
\begin{eqnarray}
&&R(r) = P^{2} - \Delta\bigl[r^{2} + (L_{z}-aE)^{2} + Q\bigr], \nonumber \\
&&\Theta_{\zeta}(\zeta) = Q(1-\zeta^{2}) - \zeta^{2}\bigl[\beta(1-\zeta^{2})+L_{z}^{2}\bigr], \nonumber \\
&&T_{r} = \frac{(r^{2}+a^{2})P}{\Delta} - a^{2}E + aL_{z},\quad \Phi_{r} = \frac{aP}{\Delta} - aE, \nonumber \\
&&T_{\zeta} = a^{2} E \zeta^{2},\quad \Phi_{\zeta}=\frac{L_{z}}{1-\zeta^{2}}.
\label{eq:R}
\end{eqnarray}
The radial and polar motions manifest as independent oscillations that additively drive the coordinate time $t$ and the azimuthal angle $\varphi$. For stable bound orbits, $R(r)$ admits four real roots, $r_{1}>r_{2}>r_{3}\ge r_{4}\ge0$, confining the radial motion to $[r_{2},r_{1}]$. Parameterizing the turning points with the semi-latus rectum $p$ and eccentricity $e$ gives $r_{1}=p/(1-e)$ and $r_{2}=p/(1+e)$. The polar potential restricts the motion to $\zeta^{2} \le \zeta_{-}^{2} \le 1$. The orbital inclination $\iota$ is defined by the maximum latitude attained by the trajectory~\cite{Schmidt:2002qk,Katz:2021yft}
\begin{equation}
x\equiv\cos\iota=\operatorname{sgn}(L_{z})\sqrt{1-\zeta_{-}^{2}}, \qquad 0\le\iota\le\pi .
\label{eq:xinc}
\end{equation}
The inclination of the angular momentum vector, $\cos\iota_{L}=L_{z}/\sqrt{L_{z}^{2}+Q}$~\cite{Ryan:1995xi,Hughes:1999bq}, differs from $\iota$ by a quadrupolar effect of $\mathcal{O}(a^{2}(1-E^{2})/L_{z}^{2})$, which remains below $1.2^{\circ}$ for near separatrix orbits. Eliminating $Q$ recasts the radial potential as a quadratic polynomial in $(E,L_{z})$
\begin{equation}
R=f(r)E^{2}-2g(r)EL_{z}-h(r)L_{z}^{2}-d(r),
\label{eq:schmidt}
\end{equation}
where $f=r^{4}+a^{2}[r(r+2)+\zeta_{-}^{2}\Delta]$, $g=2ar$, $h=r(r-2)+\zeta_{-}^{2}\Delta/x^{2}$, and $d=\Delta(r^{2}+a^{2}\zeta_{-}^{2})$. The turning-point conditions $R(r_{1})=R(r_{2})=0$ are linear in $E^{2}$, $EL_{z}$, and $L_{z}^{2}$~\cite{Wilkins:1972rs,Fujita:2009bp}. With the notation $[uv]\equiv u(r_{1})v(r_{2})-u(r_{2})v(r_{1})$, solving these constraints yields algebraic expressions for the energy and angular momentum
\begin{eqnarray}
&&E^{2}=\frac{[dh][fh]+2[dg][gh]\mp2\sqrt{\mathcal{D}}}{[fh]^{2}+4[fg][gh]},  \nonumber \\
&&\mathcal{D}=[gh]\bigl([gh][dg]^{2}+[fh][dg][dh]-[fg][dh]^{2}\bigr),    \nonumber \\
&&L_{z}=\frac{-g_{1}E\pm\sqrt{g_{1}^{2}E^{2}+h_{1}\bigl(f_{1}E^{2}-d_{1}\bigr)}}{h_{1}},
\label{eq:Lz}
\end{eqnarray}
where $u_{1}\equiv u(r_{1})$ and the upper signs correspond to prograde motion.

\par
\begin{figure*}[htbp]
\centering
\includegraphics[width=14cm,height=10cm]{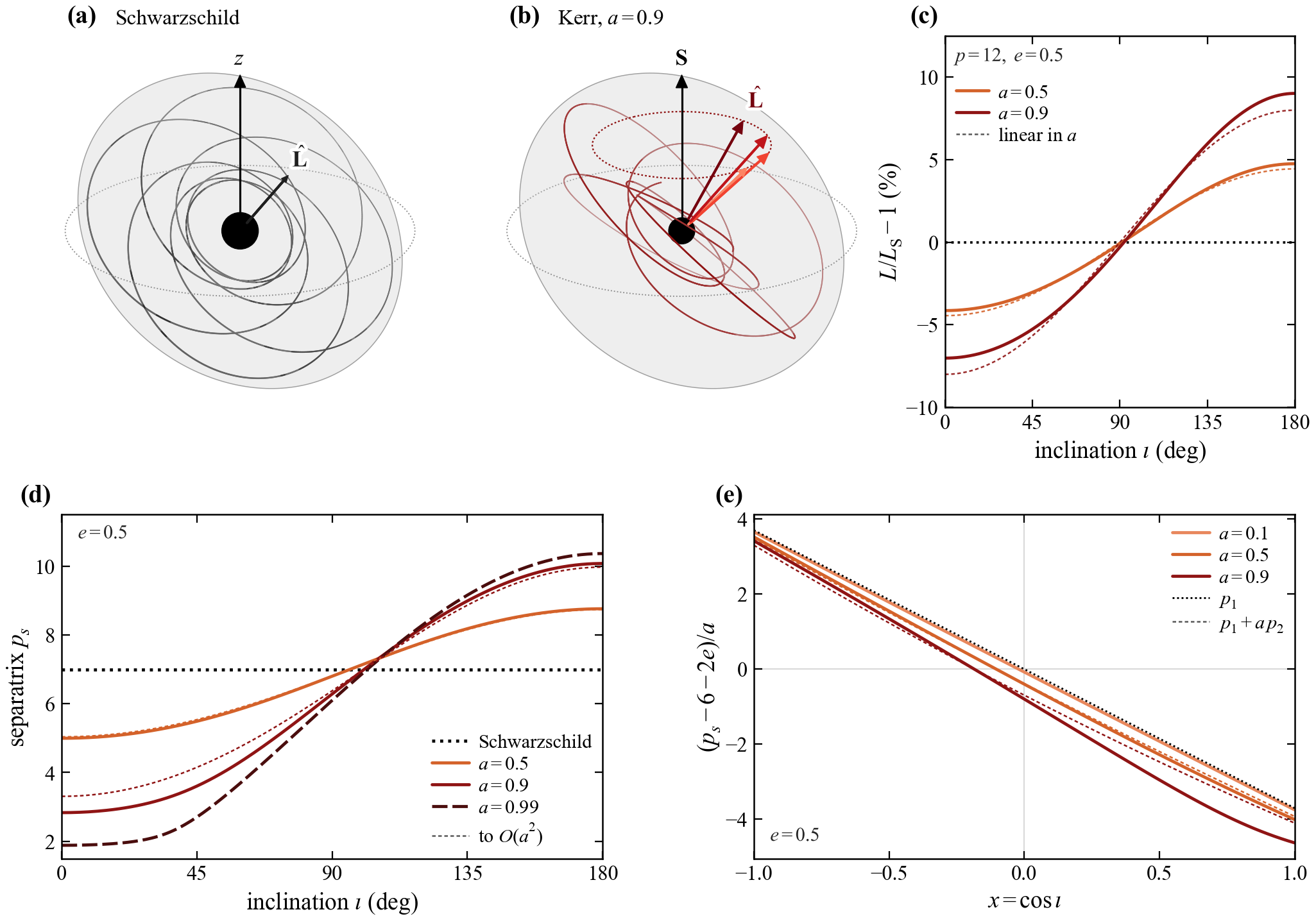}
\caption{Bound geodesics and stability boundaries.
(a) A planar orbit in the Schwarzschild spacetime.
(b) The corresponding orbit in the Kerr spacetime ($a=0.9$), showing nodal precession driven by frame dragging.
(c) Total angular momentum shift relative to the Schwarzschild baseline as a function of inclination.
(d) Separatrix location $p_{s}$ versus inclination at $e=0.5$.
(e) Spin-normalized separatrix shift, showing the leading-order dependence on $x=\cos\iota$.}
\label{fig:geodesics}
\end{figure*}

In the non-rotating limit ($a=0$), the system collapses to the equatorial formulation
\begin{equation}
E^{2}=\frac{(p-2)^{2}-4e^{2}}{p\,(p-3-e^{2})},\qquad
L^{2}=\frac{p^{2}}{p-3-e^{2}}.
\label{eq:schw}
\end{equation}
Due to spherical symmetry, any inclined Schwarzschild orbit maps to the equatorial plane via a global rotation [Fig.~\ref{fig:geodesics}(a)]. In the Kerr spacetime, frame dragging causes the orbital plane normal to precess around the spin axis [Fig.~\ref{fig:geodesics}(b)]. Consequently, at fixed $(p,e)$, the dynamics depend on $\iota$. Expanding the radial potential shows that the scalar spin--orbit coupling enters at linear order as $-4aEL_{z}r\propto\boldsymbol{S}\cdot\boldsymbol{L}$
\begin{equation}
\frac{\partial E}{\partial a}\Big|_{a=0}=-\frac{x\,(1-e^{2})^{2}}{p\,(p-3-e^{2})^{3/2}},\quad
\frac{\partial L}{\partial a}\Big|_{a=0}=-\frac{x\,(3+e^{2})\,E_{\rm S}}{p-3-e^{2}},
\label{eq:dEda}
\end{equation}
where $E_{\rm S}$ is the Schwarzschild energy. To leading order, the radial motion depends on inclination through the projected spin $a\cos\iota$ [Fig.~\ref{fig:geodesics}(c)].

\par
Frame dragging introduces a second vectorial effect. The azimuthal drift corresponds to the angular velocity of locally non-rotating frames~\cite{Bardeen:1972fi}
\begin{equation}
\Bigl(\frac{\mathrm{d}\varphi}{\mathrm{d}t}\Bigr)_{\rm drag} = \frac{\Phi_{r}}{T_{r}+T_{\zeta}}=\frac{2a}{r^{3}}+\mathcal{O}(a^{2}).
\label{eq:drag}
\end{equation}
This dragging rate is independent of $L_{z}$ and $\theta$. It coherently rotates the orbital structure and drives the macroscopic nodal precession. The separatrix $p_{s}(a,e,x)$ bounds the domain of stable orbits, defined by the homoclinic limit $r_{2}=r_{3}$~\cite{Glampedakis:2002ya,Levin:2008yp}. Unlike the static case ($p_{s}=6+2e$), the Kerr separatrix depends on inclination [Fig.~\ref{fig:geodesics}(d)]. Perturbative expansion in $a$ yields
\begin{eqnarray}
&&p_{s}=6+2e+a\,p_{1}(e,x)+a^{2}p_{2}(e,x)+\mathcal{O}(a^{3}), \nonumber \\
&&p_{1}=-4\sqrt{2}\,x\sqrt{\frac{1+e}{3+e}},  \nonumber  \\
&&p_{2}=-\frac{e^{3}+9e^{2}+11e+11-4(2e^{2}+e+1)\,x^{2}}{2(3+e)^{2}} .
\label{eq:psep}
\end{eqnarray}
The linear shift is governed by $a\cos\iota$, while the isolated inclination effect enters through the $x^{2}$ term in $p_{2}$ [Fig.~\ref{fig:geodesics}(e)]. An orbit with fixed $(p,e)$ is dynamically stable only below a maximum critical inclination $\iota_{\max}(p)$.

\par
Parameterizing the trajectory in Mino time yields the fundamental periods of the radial and polar motions: $\Lambda_{r}=2\int_{r_{2}}^{r_{1}} R^{-1/2}\,\mathrm{d}r$ and $\Lambda_{\theta}=4\int_{0}^{\zeta_{-}} \Theta_{\zeta}^{-1/2}\,\mathrm{d}\zeta$. A distant observer measures three fundamental frequencies~\cite{Schmidt:2002qk,Drasco:2003ky}
\begin{equation}
\Omega_{r}=\frac{\Upsilon_{r}}{\Gamma},\qquad
\Omega_{\theta}=\frac{\Upsilon_{\theta}}{\Gamma},\qquad
\Omega_{\varphi}=\frac{\Upsilon_{\varphi}}{\Gamma},
\label{eq:freqs}
\end{equation}
where $\Upsilon_{i}=2\pi/\Lambda_{i}$ and $\Gamma$ is the orbit-averaged time-dilation factor. Any dynamic quantity evaluated along the orbit decomposes into a Fourier series of harmonics $m\Omega_{\varphi}+j\Omega_{\theta}+n\Omega_{r}$. Using partial fraction decomposition, the radial averages reduce to closed forms expressed via complete elliptic integrals $\Pi(n|k)$~\cite{Byrd:1971bey}
\begin{eqnarray}
&&\langle r\rangle_{r}=r_{3}+(r_{2}-r_{3})\,\frac{\Pi_{1}}{\mathsf{K}}, \nonumber \\
&&\langle r^{2}\rangle_{r}=r_{3}^{2}+2r_{3}(r_{2}-r_{3})\,\frac{\Pi_{1}}{\mathsf{K}} + (r_{2}-r_{3})^{2}\,\frac{\Pi_{2}}{\mathsf{K}},\nonumber \\
&&\Bigl\langle\frac{1}{r-c}\Bigr\rangle_{r} = \frac{1}{r_{3}-c} + \Bigl(\frac{1}{r_{2}-c}-\frac{1}{r_{3}-c}\Bigr)\frac{\Pi(n_{c}|k_{r})}{\mathsf{K}},
\label{eq:radavg}
\end{eqnarray}
where $c=r_{\pm}$, $\mathsf{K}$ is the complete elliptic integral of the first kind, $\Pi_{1}=\Pi(n_{r}|k_{r})$, and $\Pi_{2}$ is a linear combination of elliptic integrals~\cite{Byrd:1971bey}. These closed forms preserve precision at near-separatrix proximities ($p-p_{s}=10^{-6}$), avoiding the truncation errors of numerical quadratures.

\par
\begin{figure*}[htbp]
  \centering
  \includegraphics[width=14cm,height=10cm]{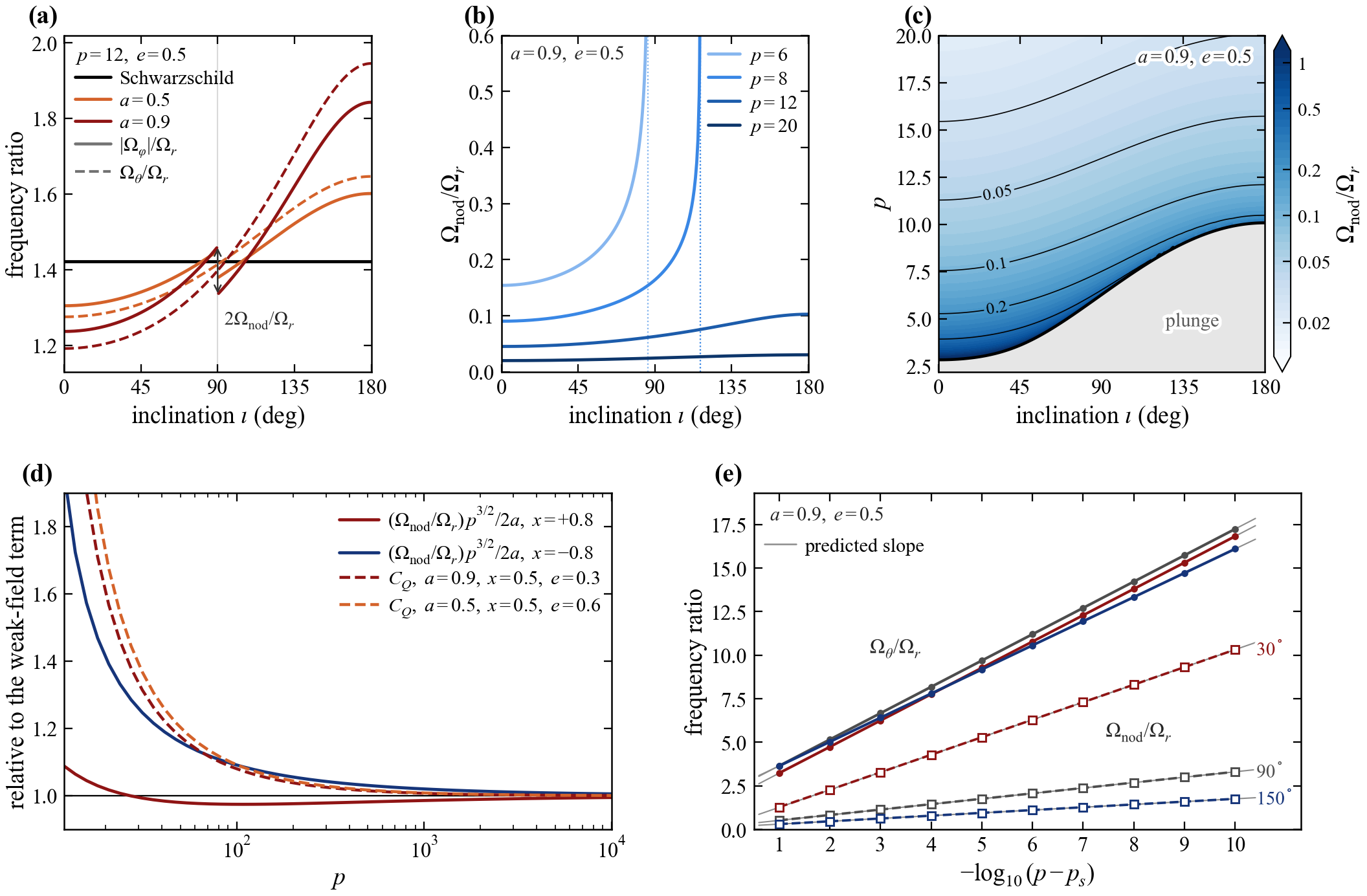}
  \caption{Fundamental frequencies and nodal precession.
  (a) Azimuthal ($|\Omega_{\varphi}|/\Omega_{r}$) and polar ($\Omega_{\theta}/\Omega_{r}$) frequency ratios.
  (b) Nodal precession ratio $\nu=\Omega_{\rm nod}/\Omega_{r}$ versus inclination. The dotted lines denote the plunge boundary $\iota_{\max}(p)$.
  (c) Contour map of $\nu$ in the $(\iota,p)$ plane.
  (d) Weak-field asymptotic behavior, isolating the Lense-Thirring dipole and mass-quadrupole components.
  (e) Logarithmic divergence of the frequency ratios $\Omega_{\theta}/\Omega_{r}$ and $\nu=\Omega_{\rm nod}/\Omega_{r}$ near the separatrix. Grey lines denote analytical predictions.}  
  \label{fig:frequencies}
\end{figure*}

Physically, $\Omega_{\theta}$ characterizes the in-plane frequency, while the residual azimuthal motion defines the uniform rotation of the orbital plane around the spin axis. This gives the nodal precession frequency~\cite{Stella:1997tc,Ingram:2009vm}
\begin{equation}
\Omega_{\rm nod}\equiv\Omega_{\varphi}-s\,\Omega_{\theta},\qquad
s\equiv\operatorname{sgn}L_{z}.
\label{eq:nod}
\end{equation}
We define the dimensionless nodal advance per radial period as $\nu\equiv\Omega_{\rm nod}/\Omega_{r}$ [Fig.~\ref{fig:frequencies}(a)].

\par
For a static, spherically symmetric spacetime ($\mathrm{d}s^{2}=-F(r)\mathrm{d}t^{2}+H(r)\mathrm{d}r^{2}+r^{2}\mathrm{d}\Omega^{2}$), the angular equations yield
\begin{equation}
\Bigl(\frac{\mathrm{d}\zeta}{\mathrm{d}\lambda}\Bigr)^{2}=L^{2}(1-\zeta^{2})-L_{z}^{2},
\qquad
\frac{\mathrm{d}\varphi}{\mathrm{d}\lambda}=\frac{L_{z}}{1-\zeta^{2}},
\label{eq:sphang}
\end{equation}
which decouple from the radial coordinate $r$. The azimuthal phase accumulated over one polar period integrates to $2\pi\operatorname{sgn}L_{z}$. Therefore, regardless of the radial metric profile,
\begin{equation}
\Omega_{\theta}=|\Omega_{\varphi}|,\qquad \Omega_{\rm nod}=0 .
\label{eq:sphthm}
\end{equation}
A non-zero $\Omega_{\rm nod}$ thus directly indicates a departure from spherical symmetry. In the Kerr metric, frame dragging ensures $\nu>0$ across the parameter space of stable bound orbits [Figs.~\ref{fig:frequencies}(b) and \ref{fig:frequencies}(c)]. In the weak-field limit, normalizing the nodal regression by the Keplerian mean motion $\Omega_{r}\simeq (p/(1-e^{2}))^{-3/2}$ yields
\begin{eqnarray}
&&\nu = \frac{2a}{p^{3/2}}-\frac{3}{2}\,\frac{a^{2}x}{p^{2}}+\dots, \nonumber \\
&&\frac{\Omega_{\theta}}{\Omega_{r}}-1 = \frac{3}{p}-\frac{6ax}{p^{3/2}}+\dots,  \nonumber \\
&&\frac{|\Omega_{\varphi}|}{\Omega_{r}}-1 = \frac{3}{p}+\frac{2a(s-3x)}{p^{3/2}}+\dots.
\label{eq:weak}
\end{eqnarray}
The leading-order term corresponds to the Lense--Thirring dipole effect. The subleading term scales with the mass quadrupole $M_{2}$ modulated by $\cos\iota$. Because inclination breaks the degeneracy between the two, the observable
\begin{equation}
C_{Q}(p)\equiv\frac{\nu(a,x)-\nu(a,-x)}{-3a^{2}x/p^{2}}
\;\xrightarrow{\;p\to\infty\;}\;1
\label{eq:CQ}
\end{equation}
isolates the quadrupole deformation [Fig.~\ref{fig:frequencies}(d)].

\par
As the orbit approaches the separatrix ($p \to p_{s}$), it enters the strong-field zoom--whirl regime. The radial period diverges logarithmically, governed by the Lyapunov exponent $\kappa$ of the unstable spherical orbit
\begin{equation}
\kappa^{2}=\tfrac{1}{2}R''(r_{\rm us})=(1-E^{2})(r_{1}-r_{\rm us})(r_{\rm us}-r_{4}).
\label{eq:lyap}
\end{equation}
The linear scaling $r_{2}-r_{3}\propto p-p_{s}$ provides the asymptotic divergence profiles
\begin{eqnarray}
&&\frac{\Omega_{\theta}}{\Omega_{r}} \simeq \frac{\Upsilon_{\theta}}{\pi\kappa} \ln\frac{1}{p-p_{s}}+C_{\theta}, \nonumber \\
&&\nu \simeq \frac{\Upsilon_{\rm nod}}{\pi\kappa}\ln\frac{1}{p-p_{s}}+C_{\rm nod}, \nonumber \\
&&\Upsilon_{\rm nod} = \Phi_{r}(r_{\rm us}) + L_{z}\Bigl\langle\frac{1}{1-\zeta^{2}}\Bigr\rangle_{\theta}-s\Upsilon_{\theta}.
\label{eq:loglaw}
\end{eqnarray}
Equation~\eqref{eq:loglaw} generalizes the logarithmic divergence of equatorial zoom--whirl orbits~\cite{Glampedakis:2002ya,Pretorius:2007jn} to arbitrary inclinations [Fig.~\ref{fig:frequencies}(e)]. Therefore, $\Omega_{\rm nod}$ vanishes identically under spherical symmetry and reproduces the classical Lense--Thirring rate in the far field. Near the separatrix, the secular nodal advance per radial period, $\Omega_{\rm nod}T_{r}=2\pi\nu$, grows logarithmically.

\subsection{Topological classification of $r$--$\theta$ orbital resonances}
\label{sec:periodic}
\par
A single frequency ratio classifies equatorial periodic orbits~\cite{Levin:2008mq}. An inclined orbit requires two:
\begin{equation}
q_{\theta}\equiv\frac{\Omega_{\theta}}{\Omega_{r}}-1,\qquad
q_{\varphi}\equiv\frac{|\Omega_{\varphi}|}{\Omega_{r}}-1=q_{\theta}+s\,\nu.
\label{eq:ratios}
\end{equation}
These ratios are degenerate in the static limit ($a=0$) but split in the Kerr spacetime, yielding distinct closure conditions. The condition $q_{\theta}\in\mathbb{Q}$ defines an $r$--$\theta$ periodic orbit that closes within a precessing plane~\cite{Grossman:2011ps}, corresponding to $r$--$\theta$ inspiral resonances~\cite{Flanagan:2010cd,Ruangsri:2013hra,Brink:2015roa}. The condition $q_{\varphi}\in\mathbb{Q}$ defines a $\varphi$--$r$ periodic orbit. Orbits satisfying both constraints simultaneously are fully closed.

\par
\begin{figure*}[htbp]
  \centering
  \includegraphics[width=15cm,height=10cm]{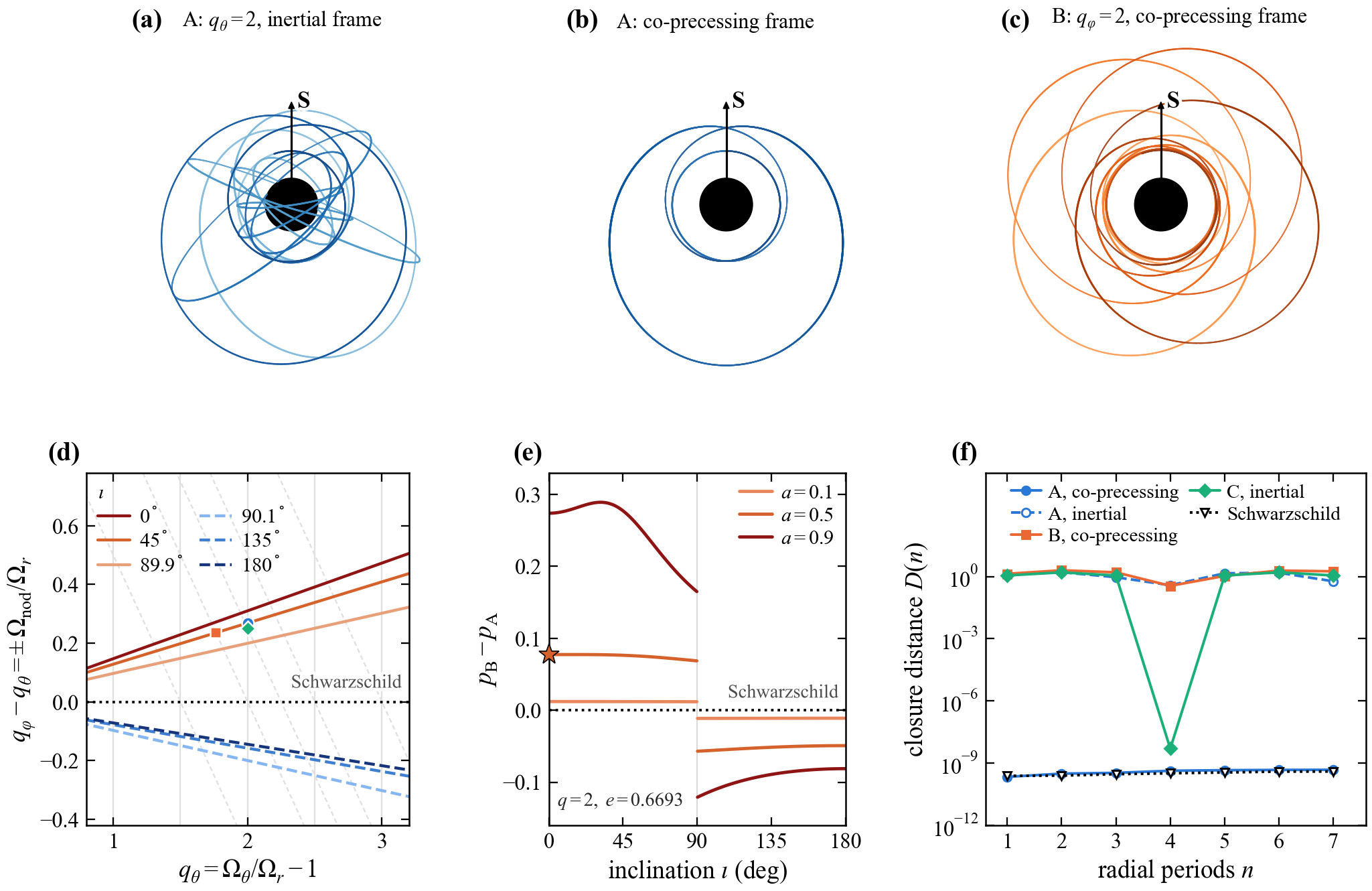}
  \caption{Notions of orbital closure at $a=0.5$ and $e=0.5$. (a)--(c) Face-on traces of $\iota=45^{\circ}$ orbits over four radial periods: (a) $r$--$\theta$ periodic $(1,2,0)$ orbit in the inertial frame, (b) the same orbit in the co-precessing frame, and (c) the $\varphi$--$r$ periodic orbit in the co-precessing frame. (d) Frequency ratios versus $p$. (e) Radial separation between the two families near the equator ($e=0.6693$). (f) Exact spatial recurrence for the fully closed $4{:}12{:}13$ orbit.}
  \label{fig:periodic}
\end{figure*}

Spatial periodicity for inclined trajectories requires $r$--$\theta$ closure. An orbit is $r$--$\theta$ periodic if and only if it closes in a frame rotating uniformly about the spin axis at the nodal rate $\Omega_{\rm nod}$~\cite{Grossman:2011ps}. Let $q_{\theta}=w+v/z$ with integers $z\ge1$, $w\ge0$, $0\le v<z$, and $\gcd(v,z)=1$. The closure period in Mino time is $\Lambda_{c}=z\Lambda_{r}=(z+zw+v)\Lambda_{\theta}$. Introducing the co-precessing azimuth $\tilde\varphi\equiv\varphi-\Omega_{\rm nod}t$ and using the relation $\Omega_{\rm nod}\Gamma=\Upsilon_{\varphi}-s\Upsilon_{\theta}$, we have
\begin{eqnarray}
\tilde\varphi = s \Upsilon_{\theta} \lambda + \bigl[\Delta\varphi_{r} - \Omega_{\rm nod}\Delta t_{r}\bigr](\eta_{r}) + \bigl[\Delta\varphi_{\theta}-\Omega_{\rm nod}\Delta t_{\theta}\bigr](\eta_{\theta}) .
\label{eq:corot}
\end{eqnarray}
Over $\Lambda_{c}$, both bracketed phases advance by integer multiples of $2\pi$. Therefore, $(r,\theta,\tilde\varphi)$ are periodic in $t$ with period $T_{c}=\Gamma\Lambda_{c}$, while $\tilde\varphi$ advances by $2\pi s(z+zw+v)$ [Figs.~\ref{fig:periodic}(a) and \ref{fig:periodic}(b)]. Full spatial closure imposes the additional requirement that $\Omega_{\rm nod}T_{c}/2\pi$ is rational.

\par
An $r$--$\theta$ closure converts each azimuthal component of the source into a discrete frequency comb. Any orbital function $S_{m}=g(r,\theta)\,e^{im\varphi}$ factorizes into $e^{im\Omega_{\rm nod}t}$ and a $T_{c}$-periodic core $g\,e^{im\tilde\varphi}$, i.e.,
\begin{eqnarray}
S_{m}(t)=\sum_{k\in\mathbb{Z}}c_{mk}\,e^{2\pi i(kf_{c}+mf_{\rm nod})t}, \qquad f_{c}\equiv\frac{1}{T_{c}}=\frac{\Omega_{r}}{2\pi z}, \qquad f_{\rm nod}\equiv\frac{\Omega_{\rm nod}}{2\pi}.
\label{eq:comb}
\end{eqnarray}
Each azimuthal mode of a source multipole manifests as a frequency comb with spacing $f_{c}$, shifted uniformly by $mf_{\rm nod}$. The generic harmonics exhibit this structure directly:
\begin{eqnarray}
m\Omega_{\varphi} + j\Omega_{\theta} + n\Omega_{r} = m \Omega_{\rm nod} + \bigl[(sm+j)(z+zw+v)+nz\bigr]\frac{\Omega_{r}}{z}.
\label{eq:harm}
\end{eqnarray}
The bracketed term is an integer, indicating that $m$ alone determines the comb. Because $\varphi$ is an ignorable coordinate, the secular evolution of the orbital constants ($E$, $L_{z}$, $Q$) depends on $r$--$\theta$ commensurabilities, while $\varphi$-resonances primarily drive recoil~\cite{Flanagan:2010cd,Flanagan:2012kg,Hirata:2010xn,vandeMeent:2014raa}. An $r$--$\theta$ closure preserves the azimuthal comb structure; a $\varphi$--$r$ commensurability entangles the frequency shift with the polar index $j$, fracturing the mode spectrum into overlapping components. This definition also maintains global continuity across polar orbits, bypassing the $2\nu$ discrete jump inherent to $q_{\varphi}$ [Figs.~\ref{fig:periodic}(d) and \ref{fig:periodic}(e)]. Hereafter, periodic orbits refer exclusively to the $r$--$\theta$ closed family, labeled as $(z,w,v)$~\cite{Levin:2008mq}. Because $q_{\theta}$ decays monotonically outward from the separatrix, each rational ratio maps to a unique trajectory~\cite{Grossman:2011ps}.

\par
The $r$--$\theta$ and $\varphi$--$r$ families remain distinct in the equatorial limit ($\iota\to0$). The $r$--$\theta$ family continuously connects to equatorial orbits commensurate with vertical epicyclic oscillations. For $e=0.6693$, the equatorial $(1,2,0)$ orbits of the two families separate radially by $\Delta p=0.012$, $0.077$, and $0.274$ for $a=0.1$, $0.5$, and $0.9$, respectively. A fully closed $4{:}12{:}13$ resonance occurs at $a=0.5$ and $e=0.5$ ($\iota=56.72^{\circ}$, $p=5.9614$) [Fig.~\ref{fig:periodic}(f)]. Exact full closure degenerates the distinct combs into a single finer comb, which remains observationally indistinguishable over finite durations.

\par
\begin{figure*}[htbp]
  \centering
  \includegraphics[width=15cm,height=10cm]{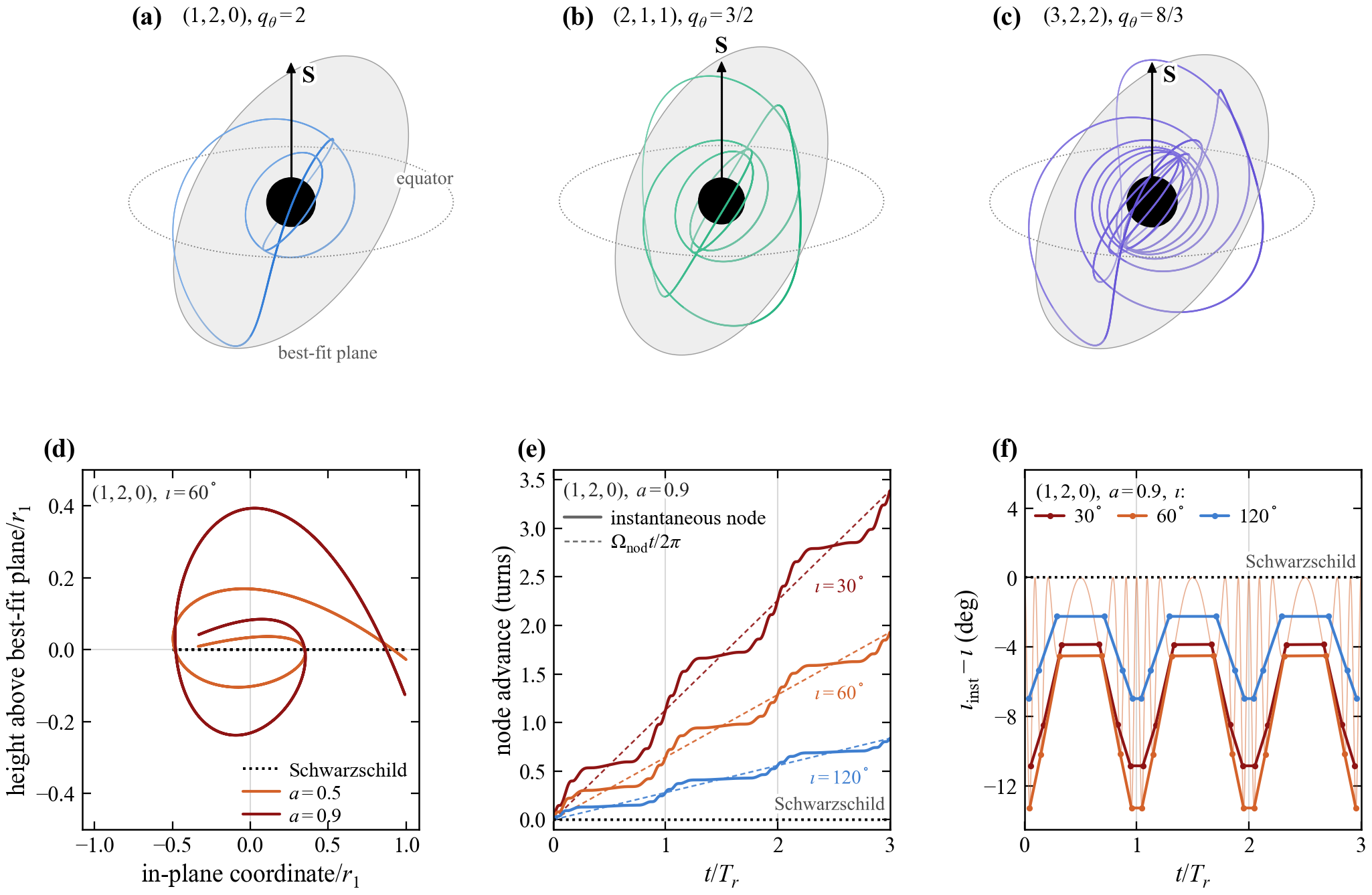}
  \caption{Strong-field orbital warping at $a=0.9$, $e=0.5$, and $\iota=60^{\circ}$. (a)--(c) The $(1,2,0)$, $(2,1,1)$, and $(3,2,2)$ orbits in the co-precessing frame, with grey disks indicating best-fit planes. (d) Edge-on projection of the $(1,2,0)$ orbit. (e) Step-like nodal advance during periastron whirls compared to uniform secular precession (dashed). (f) Instantaneous tilt relative to the baseline inclination during periastron passages.}
  \label{fig:gallery}
\end{figure*}

Figure~\ref{fig:gallery} illustrates the strong-field morphology of the $(1,2,0)$, $(2,1,1)$, and $(3,2,2)$ orbits at $a=0.9$, $e=0.5$, and $\iota=60^{\circ}$. In the co-precessing frame, each orbit closes after $z$ radial periods. Because the local frame-dragging rate scales as $r^{-3}$, the nodal advance accumulates discretely at periastron. These specific orbits spend $50\%$--$59\%$ of their coordinate time within $r<2r_{2}$, but they acquire $87\%$--$94\%$ of their total nodal shift during these brief zoom-whirl phases (compared to $34\%$ and $80\%$ for a Keplerian ellipse). The uniform secular precession thus originates from a localized strong-field effect that sets the mean slope of the azimuthal advance to $\nu$.

\par
The orbital plane dynamically tilts during these periastron whirls. Using the pseudo-flat coordinate mapping $\boldsymbol{X}=r(\sin\theta\cos\varphi,\sin\theta\sin\varphi,\cos\theta)$~\cite{Babak:2006uv}, the instantaneous orbital inclination $\iota_{\rm inst}$ is determined by the normal vector $\boldsymbol{l}=\boldsymbol{X}\times\dot{\boldsymbol{X}}$. With $A\equiv\sin^{2}\theta\,\mathrm{d}\varphi/\mathrm{d}\lambda$ and $\sin^{2}\theta\,(\mathrm{d}\theta/\mathrm{d}\lambda)^{2}=\Theta_{\zeta}$, this yields
\begin{eqnarray}
\cos\iota_{\rm inst}\equiv\frac{l_{z}}{|\boldsymbol{l}|} = \frac{\sin\theta\,A}{\sqrt{\Theta_{\zeta}(\zeta)+A^{2}}},
~~~~A=L_{z}+\sin^{2}\theta\,\frac{a\,(2Er-aL_{z})}{\Delta} .
\label{eq:tilt}
\end{eqnarray}
The inclination $\iota_{\rm inst}$ is strictly constant in the Schwarzschild spacetime. In the Kerr spacetime, the $\Delta^{-1}$ dependence of the frame-dragging term in $A$ causes $\iota_{\rm inst}$ to dip sharply at periastron, transiently aligning the orbital angular momentum toward the spin axis. For the $(1,2,0)$ orbits at $a=0.9$, this transient tilt reaches absolute deviations of $10.8^{\circ}$, $13.3^{\circ}$, and $7.0^{\circ}$ for baseline inclinations of $30^{\circ}$, $60^{\circ}$, and $120^{\circ}$, respectively.

\par
These discrete nodal jumps and transient tilts warp the orbital structure. The $(1,2,0)$ orbit at $\iota=60^{\circ}$ deviates from its best-fit geometric plane by up to $0.17\,r_{1}$ ($a=0.5$) and $0.39\,r_{1}$ ($a=0.9$). Factoring out the secular uniform precession, an inclined strong-field Kerr orbit retains a three-dimensional spatial warp localized at periastron. The resulting gravitational radiation reflects this geometry. Equation~\eqref{eq:comb} structures the spectral lines into frequency combs displaced by $mf_{\rm nod}$. The discrete jumps, tilts, and warps modulate the mode amplitudes, invalidating rigid-plane approximations. This comb structure persists until radiation reaction detunes the $q_{\theta}$ locking.

\section{Gravitational-wave emission and tests of spacetime symmetry}
\label{sec:waves}
\par
Frame dragging displaces the $m$th azimuthal comb of an $r$--$\theta$ periodic orbit by $mf_{\rm nod}$. This shift vanishes for any static, spherically symmetric metric. We adopt $M=10^{6}M_{\odot}$, $\mu=10^{-5}M$, and $D_{L}=200$~Mpc, setting the time unit to $GM/c^{3}=4.93$~s. The observer is located at $(\Theta,\Phi)=(\pi/4,\pi/4)$ relative to the spin axis. Let $S_{n}(f)$ denote the sky-averaged noise of LISA~\cite{LISA:2017pwj}, Taiji~\cite{Hu:2017mde}, or TianQin~\cite{TianQin:2015yph}. The reference source is the $(1,2,0)$ family at $e=0.6693$ and $a=0.5$. At $\iota=30^{\circ}$, this yields $p=5.612$, $\nu=0.311$, and $T_{c}=T_{r}=31.0$~min.

\subsection{Exact five-comb architecture and dynamical radiative tilt}
\label{sec:spectra}
\par
In the numerical-kludge framework, the radiation field is computed from the second time derivative of the mass quadrupole of a test body tracing a background geodesic~\cite{Thorne:1980ru,Babak:2006uv}, i.e.,
\begin{eqnarray}
h^{ij}=\frac{2\mu}{D_{L}}\bigl(a^{i}X^{j}+X^{i}a^{j}+2v^{i}v^{j}\bigr), \qquad \boldsymbol{v}=\frac{\boldsymbol{X}'}{t'}, \qquad \boldsymbol{a}=\frac{\boldsymbol{X}''t'-\boldsymbol{X}'t''}{t'^{3}},
\label{eq:quad}
\end{eqnarray}
where $\boldsymbol{X}$ is the flat-space position defined in Sec.~\ref{sec:periodic}, primes denote Mino-time derivatives, and $r''=R'/2$ alongside $\zeta''=\Theta_{\zeta}'/2$ follow analytically from Eq.~\eqref{eq:mino}. Projecting onto the transverse basis of the observer yields the polarizations $h_{+,\times}=\frac12e^{ij}_{+,\times}h_{ij}$, with $e_{+}=\hat e_{\Theta}\hat e_{\Theta}-\hat e_{\Phi}\hat e_{\Phi}$ and $e_{\times}=\hat e_{\Theta}\hat e_{\Phi}+\hat e_{\Phi}\hat e_{\Theta}$. Because Eq.~\eqref{eq:quad} is evaluated along the exact geodesic, it inherits the harmonic structure of Eq.~\eqref{eq:harm}, consistent with Teukolsky waveforms~\cite{Drasco:2005kz,Teukolsky:1973ha}. The spectral line positions are exact, while the quadrupole approximation affects only the mode amplitudes. In the equatorial limit, the waveform peaks at $\max|h_{+}|=1.90\times10^{-21}$.

\par
\begin{figure*}[htbp]
\centering
\includegraphics[width=15cm,height=10cm]{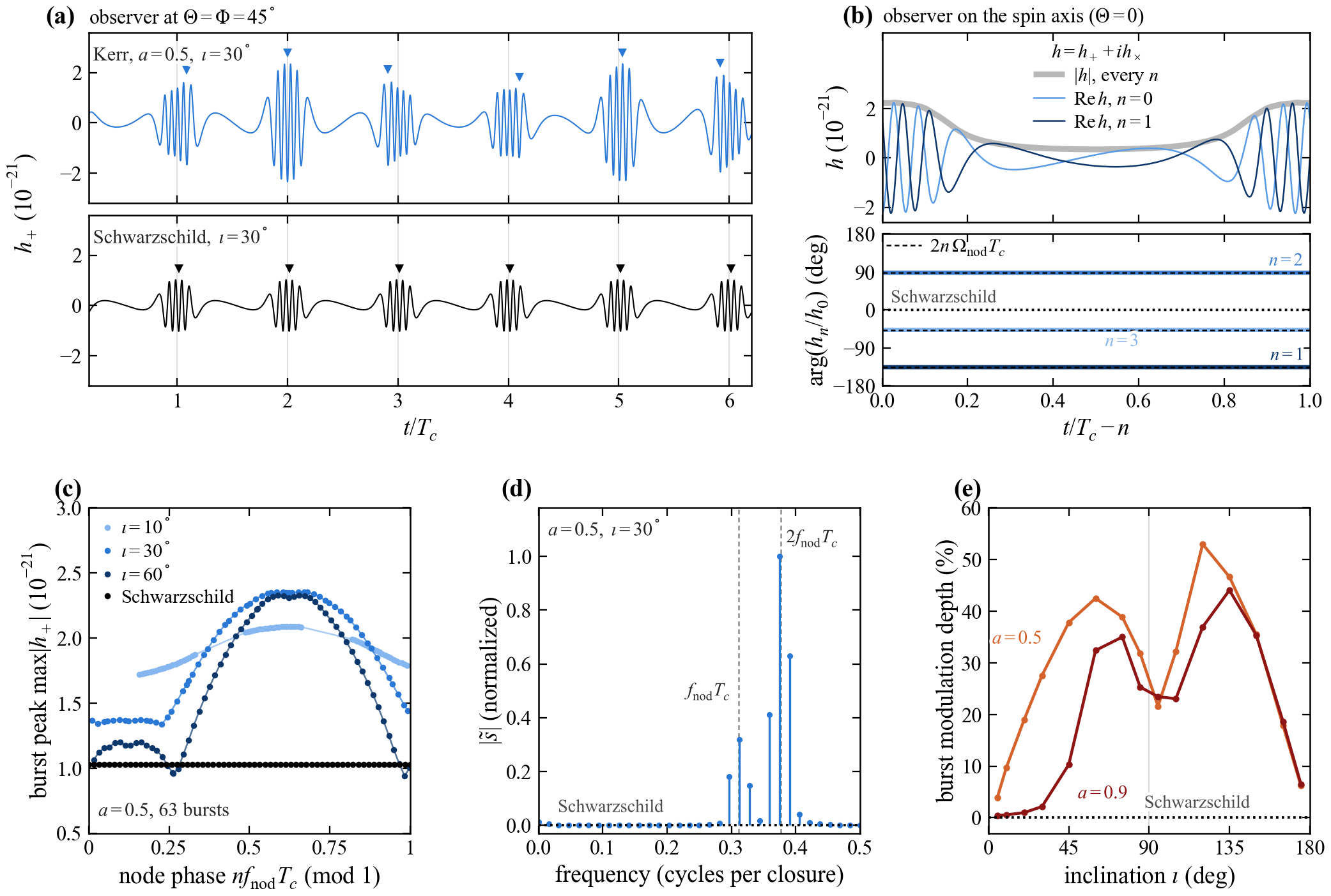}
\caption{Kludge waveforms of the $(1,2,0)$ family at $a=0.5$ and $e=0.6693$.
(a) Plus polarization over six closures at $\iota=30^{\circ}$ for an observer at $\Theta=\Phi=45^{\circ}$. Triangles mark whirl burst peaks, varying across closures only in Kerr spacetimes.
(b) On-axis complex strain. Top: the amplitude is identical across closures, while the real part shifts in phase. Bottom: the relative phase shift tracks the accumulated nodal precession (dashed).
(c) Peaks of 63 successive bursts versus fractional node phase trace a single continuous beam pattern.
(d) Spectrum of the stroboscopic sequence at $\iota=30^{\circ}$. Dashed lines mark the nodal frequency and its harmonic.
(e) Modulation depth of burst peaks versus inclination.}
\label{fig:waveform}
\end{figure*}

Nodal precession imprints a signature directly in the time domain. A spatial rotation by $\delta\varphi$ about the spin axis multiplies the azimuthal components of the strain,
\begin{eqnarray}
H_{2}=h^{xx}-h^{yy}+2ih^{xy},~~~H_{1}=h^{xz}+ih^{yz},~~~H_{0}=h^{zz}-\tfrac{1}{2}\bigl(h^{xx}+h^{yy}\bigr),
\label{eq:Hm}
\end{eqnarray}
by $e^{im\,\delta\varphi}$. Following Eq.~\eqref{eq:comb}, one closure of an $r$--$\theta$ periodic orbit corresponds to a rotation by $\Omega_{\rm nod}T_{c}$:
\begin{equation}
H_{m}(t+T_{c})=e^{im\Omega_{\rm nod}T_{c}}\,H_{m}(t) .
\label{eq:Hshift}
\end{equation}
An inclined Schwarzschild waveform is therefore $T_{c}$-periodic, whereas a Kerr source is globally rotated by $2\pi\nu_{c}$ (with $\nu_{c}\equiv f_{\rm nod}T_{c}=z\nu$) upon each closure. The temporal whirl bursts vary across successive closures (Fig.~\ref{fig:waveform}(a)), similar to the modulation in spinning binaries~\cite{Apostolatos:1994mx}. This azimuthal symmetry isolates $f_{\rm nod}$ across three observational scales. On the spin axis, the complex strain $h_{+}+ih_{\times}=H_{2}/2$ has a $T_{c}$-periodic modulus while its phase shifts by $2\Omega_{\rm nod}T_{c}$ per closure (Fig.~\ref{fig:waveform}(b)). Moving off-axis introduces the observer's azimuth $\Phi$ via $b_{m}\propto e^{-im\Phi}$, enforcing the translation symmetry
\begin{equation}
h_{+}(t+nT_{c};\Theta,\Phi)=h_{+}(t;\Theta,\Phi-2\pi n\nu_{c}) .
\label{eq:beam}
\end{equation}
The $n$th closure burst is geometrically identical to the zeroth burst viewed from an angle shifted by $-2\pi n\nu_{c}$. Plotted sequentially, the burst peaks trace a continuous azimuthal beam pattern (Fig.~\ref{fig:waveform}(c)) with a modulation depth scaling with inclination, reaching $43\%$ at $\iota=60^{\circ}$ for $a=0.5$ (Fig.~\ref{fig:waveform}(e)). Finally, the stroboscopic sequence
\begin{equation}
s_{n}\equiv h_{+}(t_{0}+nT_{c})=\sum_{m=-2}^{2}C_{m}\,e^{2\pi i m\nu_{c}n}
\label{eq:strobe}
\end{equation}
condenses the spectrum to discrete lines at $m\nu_{c}$ (mod~1) (Fig.~\ref{fig:waveform}(d)), a resonant structure absent in spherical symmetry.

\par
\begin{figure*}[htbp]
\centering
\includegraphics[width=15cm,height=10cm]{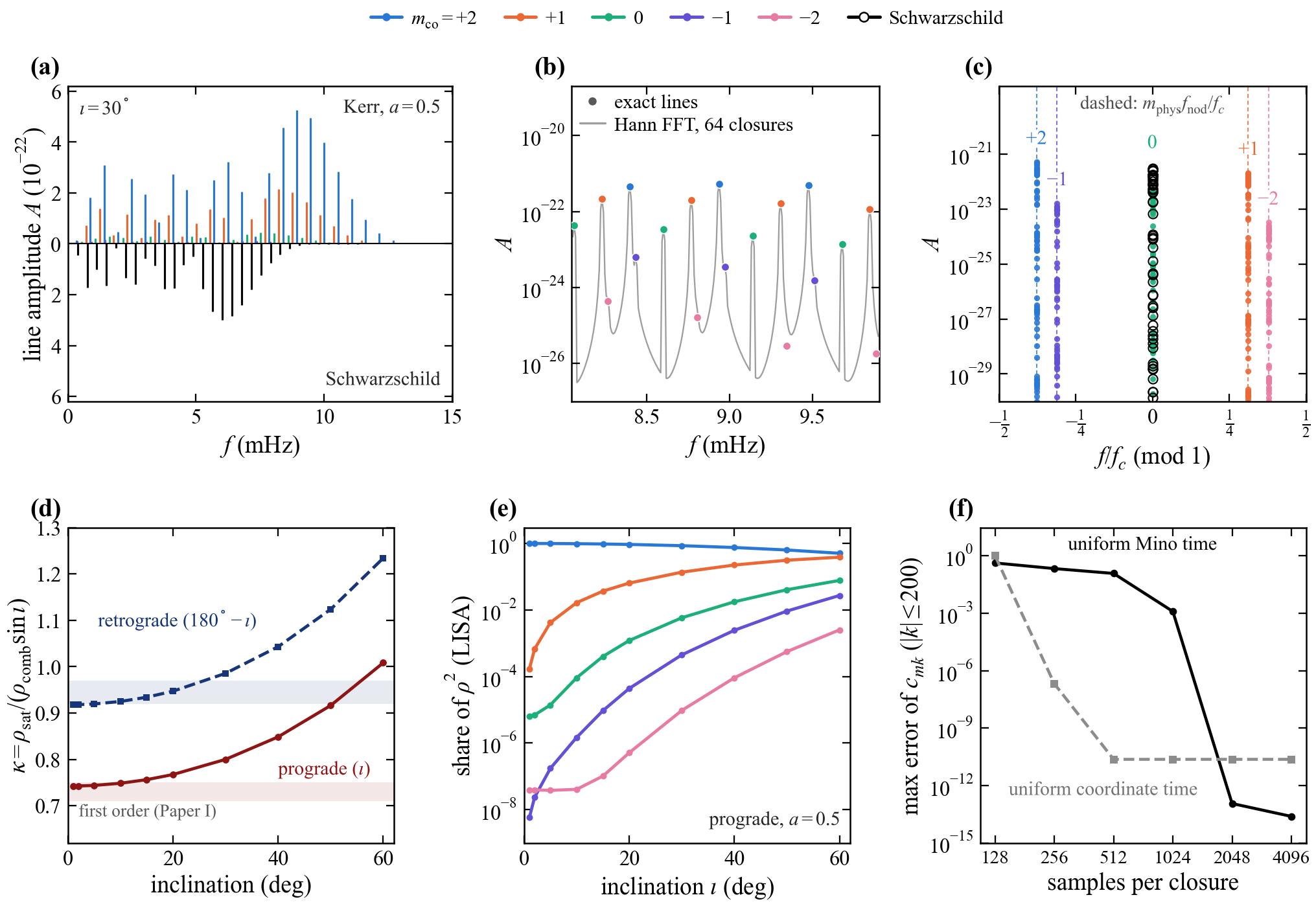}
\caption{Line spectra at $a=0.5$.
(a) Exact line amplitudes for a Kerr orbit at $\iota=30^{\circ}$ (top) and the degenerate Schwarzschild orbit (bottom, mirrored).
(b) Spectral peak lines (dots) overlaid on a Hann-windowed Fourier transform of 64 closures (grey).
(c) Spectrum folded at the main comb spacing. Kerr lines group into five columns at fractional nodal offsets (dashed).
(d) Satellite scaling coefficient in LISA for prograde (solid) and retrograde (dashed) orbits. Shaded bands are first-order analytical estimates.
(e) Distribution of total signal power across the five combs for prograde orbits.
(f) Numerical convergence of the Fourier coefficients versus sampling resolution per closure.}
\label{fig:lines}
\end{figure*}

The five constituent combs are separated in the co-precessing frame, where $H_{m}e^{-im\Omega_{\rm nod}t}$ is $T_{c}$-periodic. Its Fourier coefficients are given by integrals over a single closure in Mino time
\begin{equation}
c_{mk}=\frac{1}{T_{c}}\int_{0}^{\Lambda_{c}}H_{m}\, e^{-i(k\omega_{c}+m\Omega_{\rm nod})t(\lambda)}\,T(\lambda)\,\mathrm{d}\lambda,
\label{eq:cmk}
\end{equation}
with $\omega_{c}=2\pi f_{c}$ and $T=\mathrm{d}t/\mathrm{d}\lambda$. An evenly spaced grid of $4096$ points per closure resolves $c_{mk}$ to a precision of $10^{-13}$. The full waveform is an exact superposition of five interlaced combs at $kf_{c}+mf_{\rm nod}$~\cite{Drasco:2005kz}. At $\iota=30^{\circ}$, $1002$ lines reconstruct the waveform to an accuracy of $1.5\times10^{-10}$. In the corresponding Schwarzschild orbit, each of the $200$ dominant lines degenerates five values of $m$ (Figs.~\ref{fig:lines}). Signal detectability is quantified by the noise-weighted inner product~\cite{Finn:1992wt,Cutler:1994ys}
\begin{equation}
\langle g|h\rangle=4\,\mathrm{Re}\sum_{+,\times}\int_{0}^{\infty}
\frac{\tilde g(f)\,\tilde h^{*}(f)}{S_{n}(f)}\,\mathrm{d}f .
\label{eq:inner}
\end{equation}
For an observation time $T$, a spectral line $h=2\,\mathrm{Re}(C\,e^{2\pi if_{0}t})$ achieves $\langle h|h\rangle=4|C|^{2}T/S_{n}(f_{0})$. Because lines separated by more than $1/T$ are orthogonal, the total signal-to-noise ratio decomposes over the spectrum:
\begin{eqnarray}
\rho^{2}=\sum_{\rm lines}\frac{A^{2}\,T}{S_{n}(f)},    \qquad    A=2\bigl(|C_{+}|^{2}+|C_{\times}|^{2}\bigr)^{1/2}.
\label{eq:snrlines}
\end{eqnarray}
Defining $m_{\rm co}\equiv sm$, the primary signal resides in the main comb ($m_{\rm co}=2$), flanked by the satellite comb ($m_{\rm co}=1$). The resolution of this spectral architecture depends on the power of the satellite comb. For a rigidly tilted planar source, the multipole components scale as a rotated $m'=2$ tensor, $|h_{m}|\propto|d^{2}_{m2}(\iota)|$, dictating the baseline power ratio:
\begin{eqnarray}
\frac{P_{1}}{P_{2}}=\Bigl|\frac{d^{2}_{12}(\iota)}{d^{2}_{22}(\iota)}\Bigr|^{2} = 4\tan^{2}\frac{\iota}{2},
\qquad
\rho_{\rm sat}=\kappa\,\rho_{\rm main}\sin\iota ,
\label{eq:wigner}
\end{eqnarray}
where $\kappa$ incorporates the line-of-sight projection and noise weighting, and $\rho_{\rm main}\simeq\rho$ at small inclinations. In the $\iota\to0$ limit, $\kappa$ converges to $0.742$ (prograde) and $0.918$ (retrograde), consistent with first-order analytical estimates of $0.71$--$0.75$ and $0.92$--$0.97$. In the strong-field regime, $\kappa$ departs from these planar estimates, increasing by $8\%$ at $30^{\circ}$ and $36\%$ at $60^{\circ}$ (Fig.~\ref{fig:lines}(d)). At $30^{\circ}$, the main and satellite combs contain $86\%$ and $14\%$ of the total $\rho^{2}$, respectively (Fig.~\ref{fig:lines}(e)).

\par
\begin{figure*}[htbp]
\centering
\includegraphics[width=15cm,height=10cm]{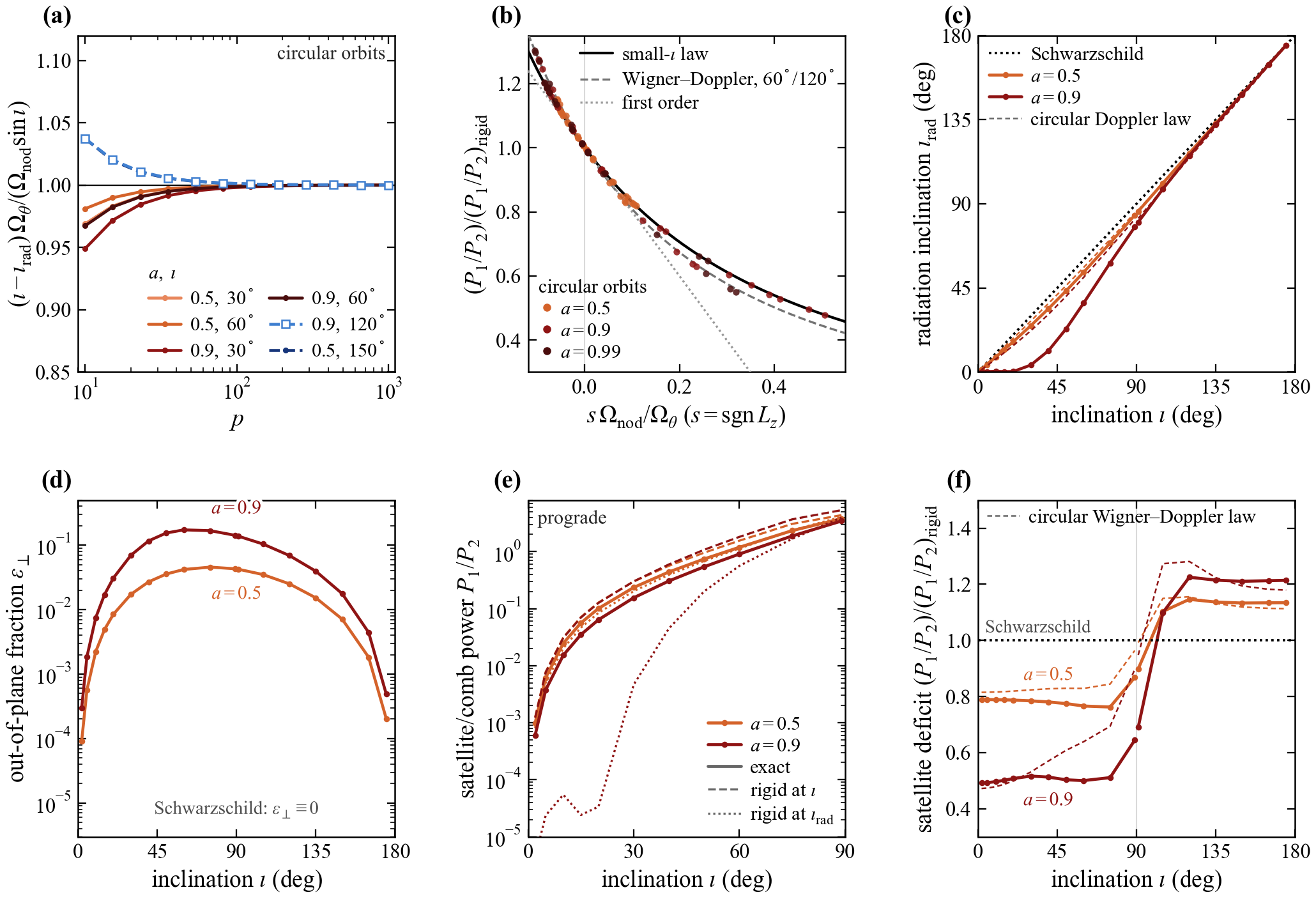}
\caption{Dynamical tilt and warp of the radiation field.
(a) Weak-field scaling of the radiation inclination offset for circular orbits.
(b) Satellite power relative to a rigid tilt for strong-field circular orbits.
(c) Radiation inclination of the $(1,2,0)$ family.
(d) Out-of-plane radiation fraction, vanishing in Schwarzschild spacetimes.
(e) Ratio of satellite to main comb power for prograde orbits, comparing exact integration (solid) with rigid projections (dashed and dotted).
(f) Satellite power of the eccentric $(1,2,0)$ family relative to a rigid tilt.}
\label{fig:radplane}
\end{figure*}

The increase in $\kappa$ originates from a dynamical tilt in the radiation field. In the inertial frame, the $m$th component of a co-precessing harmonic $\tilde\omega$ oscillates at $\omega_{m}=\tilde\omega+m\Omega_{\rm nod}$. Quadrupole radiation weights this amplitude by $\omega_{m}^{2}$. For circular orbits, $\tilde\omega=2\Omega_{\theta}$ and $|h_{m}|\propto\omega_{m}^{2}|d^{2}_{m2}(\beta)|$, with $\beta=\iota$ (prograde) or $\pi-\iota$ (retrograde). The dominant components mimic a planar source tilted at an effective radiation angle $\beta_{\rm rad}$:
\begin{eqnarray}
&&\tan\frac{\beta_{\rm rad}}{2}=\frac{\omega_{1}^{2}}{\omega_{2}^{2}}\tan\frac{\beta}{2},
\qquad \frac{\omega_{1}}{\omega_{2}}=\frac{1+s\nu_{\theta}/2}{1+s\nu_{\theta}}, \nonumber \\
&&\iota_{\rm rad}\simeq\iota-\nu_{\theta}\sin\iota,\qquad
\frac{P_{1}/P_{2}}{4\tan^{2}(\iota/2)}=\frac{\omega_{1}^{4}}{\omega_{2}^{4}}
\simeq1-2s\nu_{\theta},
\label{eq:irad}
\end{eqnarray}
where $\nu_{\theta}\equiv\Omega_{\rm nod}/\Omega_{\theta}$. The macroscopic radiation plane, defined as normal to the eigenvector of the minimum eigenvalue $\lambda_{\min}$ of $\mathsf{M}=\langle\tilde h\tilde h\rangle_{T_{c}}$ tilts toward the spin axis for both prograde and retrograde orbits. The satellite comb is suppressed for prograde trajectories and enhanced for retrograde ones. Wigner weights predict this behavior to within $2\%$ for circular orbits up to $\nu_{\theta}=0.6$ (Fig.~\ref{fig:radplane}). For zoom--whirl orbits, radiation is emitted primarily during periastron passages, where the instantaneous orbital plane dips and tilts. The emission breaks planar symmetry, yielding an out-of-plane fraction $\epsilon_{\perp}\equiv\lambda_{\min}/\mathrm{tr}\,\mathsf{M}$ up to $0.17$ at $a=0.9$. At $a=0.9$ and $\iota=30^{\circ}$, the effective inclination is $\iota_{\rm rad}=3.8^{\circ}$ (compared to the circular equivalent of $22^{\circ}$), and the rigid planar projection accounts for $3\%$ of the satellite power. The satellite comb for highly eccentric strong-field orbits is driven by the three-dimensional warp of the trajectory. Relative to a rigid tilt, the satellite power is $0.76$--$0.79$ ($a=0.5$) and $0.49$--$0.52$ ($a=0.9$) for prograde orbits below $75^{\circ}$, and increases to $1.13$--$1.15$ and $1.21$--$1.23$ beyond $120^{\circ}$. Models treating orbital orientation as a slowly varying parameter~\cite{Barack:2003fp} omit this $O(\nu_{\theta})$ amplitude modulation.

\par
\begin{figure*}[!htbp]
\centering
\includegraphics[width=15cm,height=10cm]{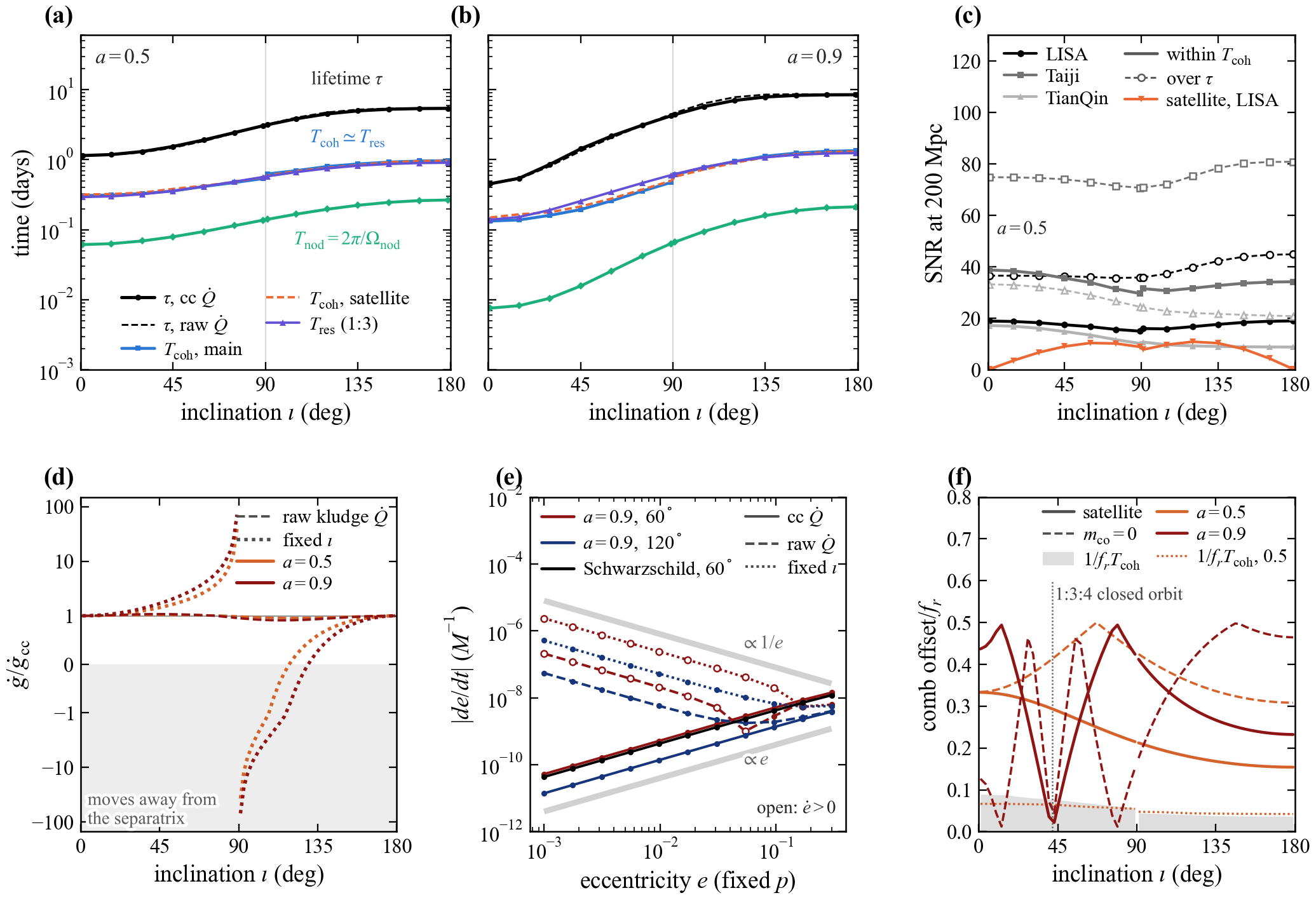}
\caption{Radiation reaction time scales along the $(1,2,0)$ family.
(a), (b) Characteristic time scales at $a=0.5$ and $0.9$: inspiral lifetime computed with symmetry-corrected (solid) and raw (dashed) Carter fluxes.
(c) Accumulated signal-to-noise ratios at $200$~Mpc ($a=0.5$).
(d) Rate of approach to the separatrix for raw and fixed-inclination models, normalized to the corrected flux.
(e) Artificial divergence of the eccentricity rate at low eccentricities when omitting the circularity correction.
(f) Frequency offsets of the satellite (solid) and central (dashed) combs relative to the resolution limit.}
\label{fig:evolution}
\end{figure*}

Radiation reaction determines the duration of this resonant structure. The fluxes of $E$ and $L_{z}$, conjugate to the spacetime Killing vectors, follow from the torus-averaged quadrupole formula:
\begin{equation}
\dot E=-\frac{\mu}{5}\bigl\langle\dddot Q_{ij}\dddot Q_{ij}\bigr\rangle,\qquad
\dot L_{z}=-\frac{2\mu}{5}\bigl\langle\epsilon_{zjk}\ddot Q_{ja}\dddot Q_{ka}\bigr\rangle,
\label{eq:fluxes}
\end{equation}
where $Q_{ij}=(X_{i}X_{j})^{\rm STF}$. Both fluxes converge to the Peters--Mathews limit as $1/p$~\cite{Peters:1963ux}. Because no conserved current anchors the Carter constant $Q$ to infinity, calculating its flux requires integrating the radiation field~\cite{Mino:2003yg}. The flat-space approximation $\dot Q=2\langle\boldsymbol{\ell}_{\perp}\cdot\dot{\boldsymbol{L}}_{\perp}\rangle$, derived via $Q\simeq L_{x}^{2}+L_{y}^{2}$, is exact in the Schwarzschild spacetime but violates the circularity preservation theorem in Kerr~\cite{Kennefick:1995za}. Expanding the Jacobian mapping,
\begin{equation}
(\dot E,\dot L_{z},\dot Q)^{\mathsf T}=\mathsf{J}\,(\dot p,\dot e,\dot x)^{\mathsf T},
\qquad
\mathsf{J}=\frac{\partial(E,L_{z},Q)}{\partial(p,e,x)} ,
\label{eq:jac}
\end{equation}
shows that because $(E,L_{z},Q)$ are even in $e$, the $e$-column of $\mathsf{J}$ vanishes as $e\to0$. A finite limit for $\dot e/e$ therefore requires $\dot Q=Q_{,p}\dot p+Q_{,x}\dot x$, which is constrained by $\dot E$ and $\dot L_{z}$. The flat-space estimate violates this constraint by $1\%$--$3\%$, resulting in $e\dot e\to{\rm const}\neq0$ as $e\to0$ [Fig.~\ref{fig:evolution}(e)]. Following Ref.~\cite{Gair:2005ih}, we enforce this symmetry
\begin{equation}
\dot Q_{\rm cc}(p,e,x)=\dot Q(p,e,x)+\dot Q_{\rm circ}(p,x)-\dot Q(p,0,x),
\label{eq:cc}
\end{equation}
where $\dot Q_{\rm circ}$ satisfies the circularity condition. These two methods bound the remaining lifetime to within $11\%$ and the local time scales to $2\%$.

\par
Physical radiation reaction produces a small $\dot\iota>0$~\cite{Ryan:1995xi}. Fixing the inclination ($\dot x=0$) alters the evolution near the separatrix. Under this constraint, Eq.~\eqref{eq:jac} reduces to $(\dot p,\dot e)$. Because $\partial L_{z}/\partial(p,e)\propto x$, the evaluated $\dot p$ and $\dot e$ scale as $1/x$, diverging for polar orbits. The fixed-inclination model~\cite{Glampedakis:2002cb} thus overestimates the plunge rate by a factor of $2.0$--$2.7$ at $60^{\circ}$ and predicts opposite signs for retrograde orbits. To resolve the near-separatrix behavior where $\det\mathsf{J}\propto g\equiv p-p_{s}$, we integrate in $\ln g$. Radiation reaction defines the coherence time $T_{\rm coh}=|\dot f|^{-1/2}$. The $1{:}3$ resonance-crossing time is $T_{\rm res}=|f_{r}\dot q_{\theta}|^{-1/2}$~\cite{Flanagan:2010cd,Ruangsri:2013hra}. Along the $(1,2,0)$ trajectory, these time scales follow the hierarchy (Figs.~\ref{fig:evolution}(a) and \ref{fig:evolution}(b)):
\begin{equation}
\underbrace{T_{\rm nod}}_{0.2\text{--}6.4\ {\rm h}}<
\underbrace{T_{\rm coh}\simeq T_{\rm res}}_{0.13\text{--}1.35\ {\rm d}}<
\underbrace{\tau}_{0.45\text{--}8.5\ {\rm d}},
\label{eq:hierarchy}
\end{equation}
with $T_{\rm res}/T_{\rm coh}=0.92$--$1.34$. Within one coherence time, the source is approximated as an $r$--$\theta$ periodic orbit, during which the orbital plane completes $3.6$ to $18$ precession cycles. The satellite and $m_{\rm co}=0$ combs are resolvable from the main comb if their frequency spacings ($\|\nu\|f_{r}$ and $\|2\nu\|f_{r}$) exceed the bin width $1/T_{\rm coh}$:
\begin{equation}
\|n\nu\|\,f_{r}T_{\rm coh}>1,\qquad n=1,2 .
\label{eq:resolve}
\end{equation}
At $a=0.5$, the satellite comb is separated by $3.6$--$5$ frequency bins. At $a=0.9$, spectral overlap occurs where $\nu=1$ ($41.85^{\circ}$) or $2\nu\in\mathbb{Z}$ ($12.40^{\circ}$ and $78.27^{\circ}$) (Fig.~\ref{fig:evolution}(f)). Over one $T_{\rm coh}$ at $a=0.5$, the total signal yields $\rho=15$--$19$ in LISA, $30$--$39$ in Taiji, and $9$--$17$ in TianQin. The satellite comb alone contributes $\rho_{\rm sat}=7$--$11$ in LISA for $30^{\circ}\le\iota\le150^{\circ}$ (Fig.~\ref{fig:evolution}(c)). Parameter extraction relies on resolving these spectral combs during the approximately locked $r$--$\theta$ phase.

\subsection{Spin resolution and the exclusion of spherical geometries}
\label{sec:analysis}
\par
Evaluating a single coherence time addresses two objectives: measuring the spin and excluding static, spherically symmetric spacetimes. Each orbit is integrated over $T_{\rm coh}$ and rounded to integer radial periods for hypothesis testing. Maximum-likelihood estimation yields the covariance $\Sigma=\Gamma^{-1}$, where $\Gamma_{ij}=\langle\partial_{i}h|\partial_{j}h\rangle$ is the Fisher matrix. For a single spectral line at SNR $\rho_{\ell}$, the matrix structure yields the scalings
\begin{equation}
\sigma_{\ln A}=\frac{1}{\rho_{\ell}},\qquad
\sigma_{f}=\frac{\sqrt{12}}{2\pi\rho_{\ell}T}\propto T^{-3/2}.
\label{eq:lineerr}
\end{equation}
The source geometry depends on four parameters $(a,p,e,x)$, but the line grid provides three fundamental frequencies. Parameter inversion thus requires amplitude information. The main comb $n_{r}f_{r}+2|f_{\varphi}|$ masks the polar frequency $f_{\theta}$ inside $|f_{\varphi}|$. The satellite comb at $n_{r}f_{r}+f_{\theta}+|f_{\varphi}|$ breaks this degeneracy and isolates $f_{\theta}$.
\begin{figure*}[htbp]
\centering
\includegraphics[width=15cm,height=10cm]{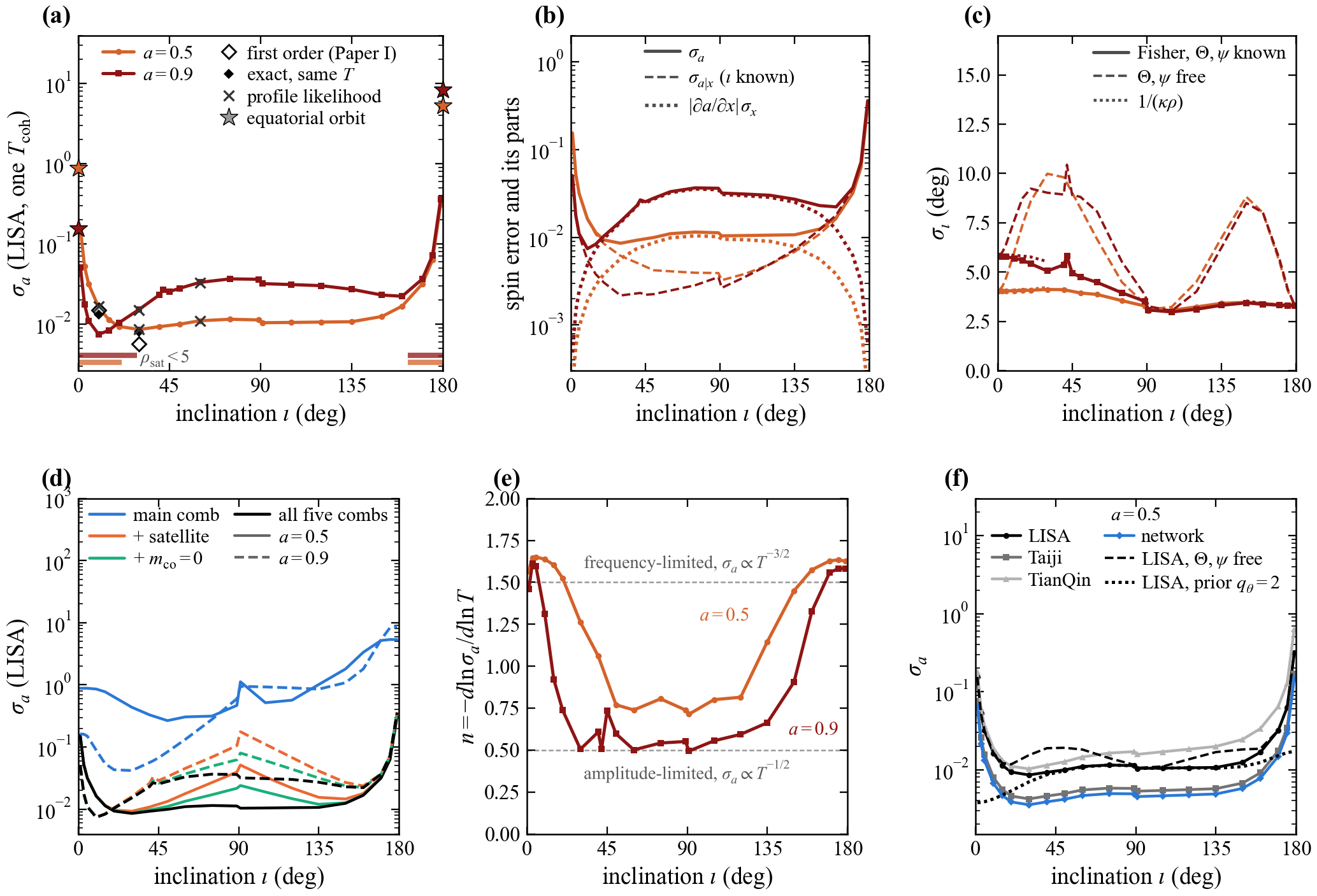}
\caption{Parameter estimation bounds for the spin over a single coherence time at $200$~Mpc.
(a) Baseline spin error $\sigma_{a}$ in LISA at $a=0.5$ and $0.9$. Stars mark equatorial references; diamonds compare frequency-limited estimates with exact models. Vertical bars mark $\rho_{\rm sat}<5$.
(b) Component decomposition of $\sigma_{a}$ (solid) via Eq.~\eqref{eq:variance}, isolating the conditional frequency error (dashed) and the amplitude degradation (dotted).
(c) Inclination error bounds for fixed (solid) and agnostic (dashed) observer positioning, with the amplitude limit $1/(\kappa\rho)$ (dotted).
(d) Spectral contributions to the spin measurement from individual and combined combs.
(e) Scaling exponent $n=-\mathrm{d}\ln\sigma_{a}/\mathrm{d}\ln T$, showing the transition from frequency-limited ($n=3/2$) to amplitude-limited ($n=1/2$) regimes.
(f) Network enhancement (LISA, Taiji, TianQin) of spin precision, alongside the penalty for an unknown sky location (dashed).}
\label{fig:fisher}
\end{figure*}

\par
This spectral decoupling divides parameter estimation into two regimes. At low inclinations, the resolvable satellite comb anchors $f_{\theta}$, placing the spin error in a frequency-limited regime with $\sigma_{a}\propto T^{-3/2}$. The inclination error is bounded by amplitude derivatives to $\sigma_{\iota}\simeq1/(\kappa\rho)$ [Fig.~\ref{fig:fisher}(c)]. At high inclinations, the orbital frequencies depend geometrically on $\iota$. Fixing the frequencies exposes a degeneracy curve in the $(a,x)$ plane with slope:
\begin{equation}
\beta_{ax}\equiv\Bigl(\frac{\partial a}{\partial x}\Bigr)_{q_{\theta},\nu,e}
=-\frac{\partial(q_{\theta},\nu)/\partial(x,p)}{\partial(q_{\theta},\nu)/\partial(a,p)}.
\label{eq:degslope}
\end{equation}
The Gaussian block-matrix identity yields:
\begin{equation}
\sigma_{a}^{2}=\sigma_{a|x}^{2}+\Bigl(\frac{\Sigma_{ax}}{\Sigma_{xx}}\Bigr)^{2}\sigma_{x}^{2},
\qquad \frac{\Sigma_{ax}}{\Sigma_{xx}}\simeq\beta_{ax}.
\label{eq:variance}
\end{equation}
The system transitions into an amplitude-limited regime where $\sigma_{a}\simeq|\beta_{ax}|\sin\iota\,\sigma_{\iota}\propto T^{-1/2}$ [Figs.~\ref{fig:fisher}(b) and \ref{fig:fisher}(e)].

\par
\begin{figure*}[htbp]
\centering
\includegraphics[width=15cm,height=10cm]{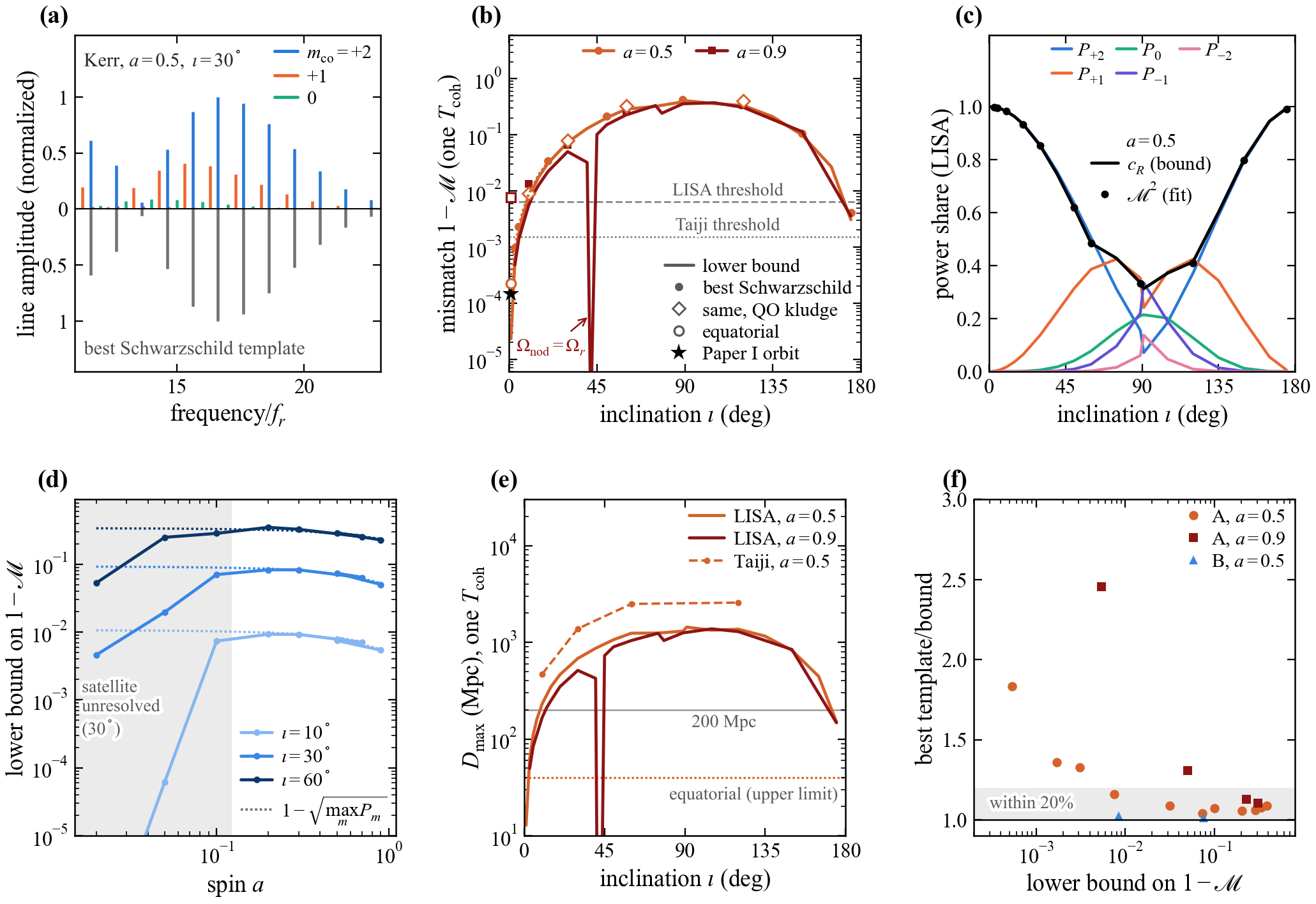}
\caption{Discrimination between Kerr and static, spherically symmetric spacetimes over one coherence time.
(a) Spectrum at $a=0.5$ and $\iota=30^{\circ}$, comparing Kerr lines (top) with the Schwarzschild template (bottom).
(b) Signal mismatch versus inclination. Lines plot the analytical lower bound; symbols denote numerical optimizations. Horizontal thresholds define distinguishability at $200$~Mpc.
(c) Fractional power across the five Kerr combs. The solid black line is the maximum captured share, tracking the numerical match $\mathcal{M}^{2}$ (dots).
(d) Minimum mismatch versus spin, converging to the single-comb limit (dotted) for $a\gtrsim0.2$.
(e) Maximum distinguishable horizon $D_{\max}$, compared to planar-orbit limits (dotted).
(f) Comparison of the analytical bound and numerical mismatch for generic periodic orbits.}
\label{fig:rotation}
\end{figure*}

These two regimes determine the parameter precision [Fig.~\ref{fig:fisher}(a)]. At $a=0.5$, $\sigma_{a}$ reaches a minimum of $0.0085$ near $\iota=30^{\circ}$ and remains below $0.032$ for $10^{\circ}\le\iota\le170^{\circ}$. At $a=0.9$, the amplitude degeneracy increases $\sigma_{a}$ to $0.037$ at high inclinations, though it remains below the equatorial baseline. Relaxing the prior on the observer's sky position inflates $\sigma_{a}$ by a maximum factor of $2.1$. A joint network (LISA, Taiji, TianQin) reduces these errors to $40\%$ of the LISA baseline [Fig.~\ref{fig:fisher}(f)]. Orbits in static, spherically symmetric spacetimes are planar
\begin{equation}
\tilde x+i\tilde y=r(t)\,e^{i\tilde\varphi(t)},\qquad
\tilde\varphi=\Omega_{\varphi}t+\Delta(t).
\label{eq:planar}
\end{equation}
The quadrupole expansion restricts the radiation to a maximum of three spectral branches
\begin{equation}
h=\sum_{k}\bigl[c_{k}e^{i(k\Omega_{r}+2\Omega_{\varphi})t}+d_{k}e^{ik\Omega_{r}t}
+c'_{k}e^{i(k\Omega_{r}-2\Omega_{\varphi})t}\bigr].
\label{eq:sphcomb}
\end{equation}
A static spacetime produces at most three combs and cannot replicate the five-comb structure of an inclined Kerr orbit. Spherical templates omit the Kerr satellite combs [Fig.~\ref{fig:rotation}(a)].

\par
An analytical lower bound on the mismatch excludes all such configurations. Defining a generalized spherical template $h_{T}$ subject to power constraints (with limits $\zeta$ and $\gamma$ for out-of-band power), the Cauchy-Schwarz inequality limits the maximum match,
\begin{equation}
\langle\hat s|h_{T}\rangle\le
\bigl[\boldsymbol{b}^{\mathsf T}\mathsf{A}^{-1}\boldsymbol{b}\,
(1+\mu\zeta^{2}+\kappa'\gamma^{2})\bigr]^{1/2},
\label{eq:dual}
\end{equation}
where $\mathsf{A}$ is the augmented Gram matrix of the projected signal. Minimizing this expression yields a lower bound for any planar-orbit quadrupole template. When the Kerr satellite combs are resolved, the bound converges to the single-comb limit
\begin{equation}
1-\mathcal{M}\gtrsim1-\Bigl(\max_{m}P_{m}\Bigr)^{1/2}
\;\xrightarrow{\;\iota\to0\;}\;\tfrac{1}{2}P_{\rm sat}.
\label{eq:onecomb}
\end{equation}

\par
At $a=0.5$, this analytical bound tracks the numerical Schwarzschild fits [Figs.~\ref{fig:rotation}(b) and \ref{fig:rotation}(c)]. When the satellites are resolved ($a\gtrsim0.2$ at $30^{\circ}$), the mismatch converges to the single-comb limit [Fig.~\ref{fig:rotation}(d)]. This mismatch yields a physical exclusion radius $D_{\max}=D_{L}(\rho/\rho_{\rm crit})$ with $\rho_{\rm crit}=[D/2(1-\mathcal{M})]^{1/2}$. Equatorial orbits limit this horizon to $\sim40$~Mpc; the inclined Kerr trajectory extends $D_{\max}$ to $230$, $680$, and $1240$~Mpc at $10^{\circ}$, $30^{\circ}$, and $60^{\circ}$, respectively, in LISA [Fig.~\ref{fig:rotation}(e)].

\par
\begin{figure*}[htbp]
\centering
\includegraphics[width=15cm,height=10cm]{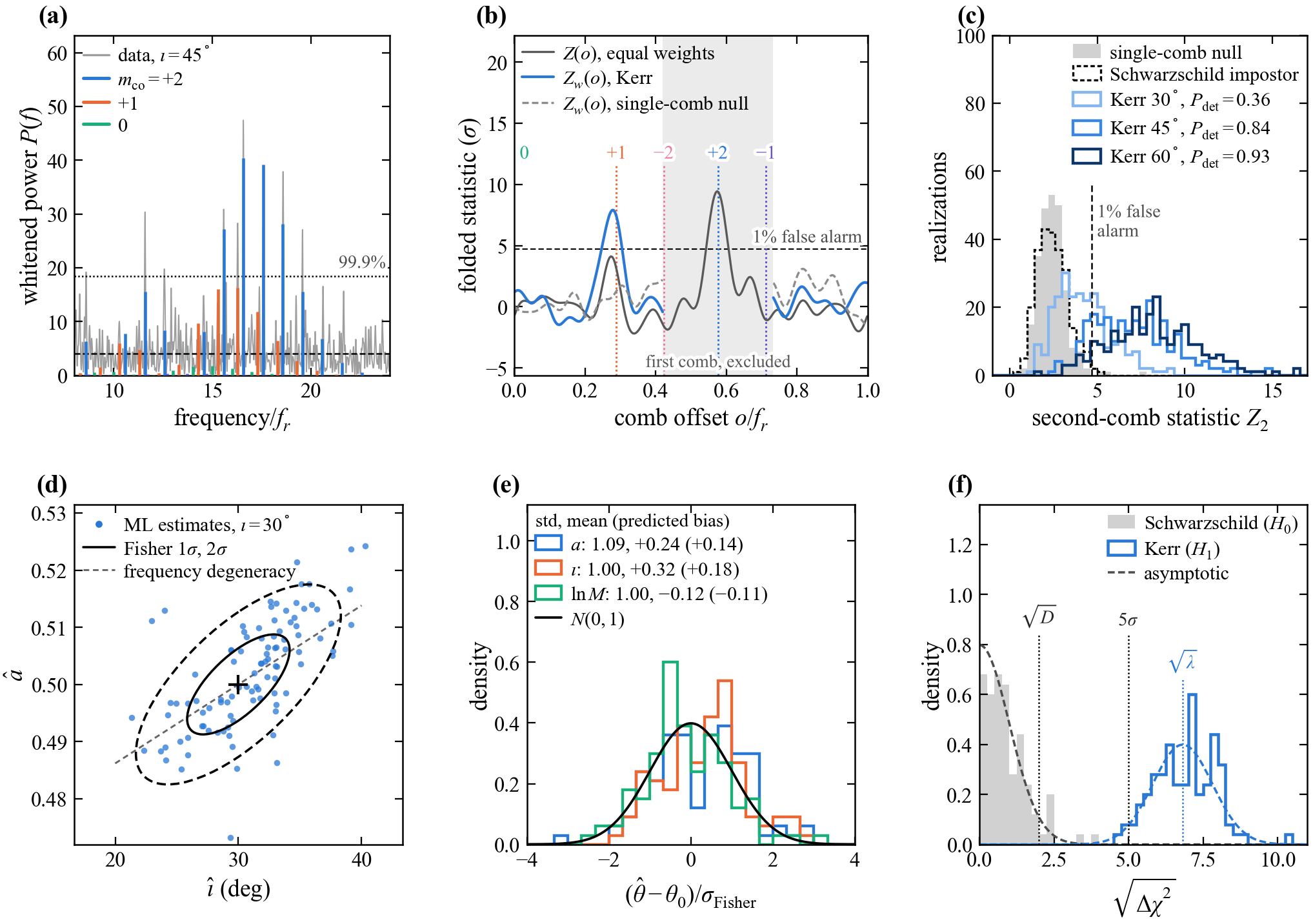}
\caption{Parameter extraction from simulated Gaussian LISA noise over one coherence time at $200$~Mpc.
(a) Whitened periodogram at $\iota=45^{\circ}$ (grey) with signal components (colored). Dashed line: $99.9\%$ noise threshold.
(b) Harmonic folding statistic. The blind scan (bottom) identifies the primary comb; the weighted scan (middle) extracts the satellite comb above the noise limit (dashed).
(c) Distribution of the second-comb statistic. Kerr injections separate from noise and the Schwarzschild template limits.
(d) Maximum-likelihood parameter scatter at $30^{\circ}$ bounded by analytical Fisher ellipses.
(e) Normalized parameter residuals matching the second-order bias predictions.
(f) Hypothesis testing via $\sqrt{\Delta\chi^{2}}$. The Kerr injection follows a noncentral $\chi^{2}$ distribution ($H_1$), separating from the Schwarzschild null hypothesis ($H_0$).}
\label{fig:simdata}
\end{figure*}

These analytical constraints are tested by injection into simulated Gaussian LISA noise. After whitening the data streams, the local power follows a $\chi^{2}_{4}$ distribution. Harmonic summing folds the spectrum across a trial comb sequence
\begin{equation}
Y(o)=\sum_{k}w_{k}\,P\bigl((k+o)f_{r}\bigr),\qquad
Z=\frac{Y-4\sum_{k}w_{k}}{\bigl(8\sum_{k}w_{k}^{2}\bigr)^{1/2}}.
\label{eq:fold}
\end{equation}
A blind scan over $(f_{r},o)$ identifies the primary comb above the false-alarm threshold [Fig.~\ref{fig:simdata}(a)]. Bootstrapping these weights for a secondary sweep extracts the displaced satellite sequence [Fig.~\ref{fig:simdata}(b)]. The satellite detection probability scales with $P_{\rm sat}$ [Fig.~\ref{fig:simdata}(c)]. Isolating the dominant phase offsets $o_{1}$ and $o_{2}$ provides an estimator $\hat\nu=\|o_{1}-o_{2}\|$ for the nodal precession rate.

\par
Parameter extraction via Gauss--Newton iterations yields a scatter consistent with the Fisher limits [Fig.~\ref{fig:simdata}(d)]. The mean residuals match the expected second-order bias $\mathbb{E}[\hat\theta^{m}-\theta^{m}_{0}]\simeq -\frac{1}{2}\Sigma^{mi}\langle\partial_{i}h|\Sigma^{jk}\partial_{j}\partial_{k}h\rangle$ [Fig.~\ref{fig:simdata}(e)]. Evaluating the model selection statistic $\Delta\chi^{2}$ orthogonal to the local Schwarzschild manifold
\begin{equation}
\Delta\chi^{2}=\lambda+2\,\boldsymbol\Delta\cdot\boldsymbol{n}
+|\mathsf{P}_{\rm K}\boldsymbol{n}|^{2}-|\mathsf{P}_{\rm S}\boldsymbol{n}|^{2},
\label{eq:lr}
\end{equation}
shows that the Kerr injection follows a noncentral $\chi^{2}$ distribution, separating from the Schwarzschild null hypothesis [Fig.~\ref{fig:simdata}(f)]. Over a single coherence time, the resolved five-comb structure enables a measurement of the nodal drift and excludes spherical geometries.

\section{Conclusions}
\label{sec:conclusions}
\par
In this work, the gravitational-wave signatures of extreme mass ratio inspirals were investigated by extending the periodic-orbit framework to inclined trajectories in the Kerr spacetime. By matching the radial and polar fundamental frequencies, a class of $r$--$\theta$ periodic orbits was isolated, which strictly close within a co-precessing spatial frame. Frame dragging induces a nodal precession that explicitly breaks the spherical symmetry of the background. Rather than accumulating uniformly, this nodal advance is predominantly generated during the strong-field periastron passages, where the orbital plane experiences transient tilts and macroscopic spatial warps.

\par
This three-dimensional geometric warping directly dictates the spectral architecture of the emitted gravitational waves. Through semi-relativistic kludge modeling, it was shown that the quadrupole radiation of an inclined periodic orbit splits into exactly five distinct frequency combs, uniformly separated by the nodal precession frequency. The strong-field dynamical tilt alters the multipole emission relative to a rigidly tilted planar source. Specifically, the power distributed into the satellite combs is suppressed for prograde orbits and enhanced for retrograde ones, an amplitude modulation entirely governed by the local frame-dragging rate at periastron.

\par
The persistence of this spectral structure was evaluated under the influence of radiation reaction. By incorporating the circularity-preserving Carter constant flux, the local orbital evolution was tracked through the strong-field regime. The hierarchy of evolutionary timescales indicates that an inspiral remains phase-locked in an approximate $r$--$\theta$ resonance over a single coherence time. During this observational window, the frequency separation between the main and satellite combs exceeds the resolvable bin width. The distinct five-comb signature is therefore preserved against radiation-reaction detuning and falls squarely within the optimal detection bands of space-borne interferometers.

\par
Detecting this fully resolved comb structure effectively breaks the parameter degeneracies inherent to equatorial orbits. In the absence of a satellite comb, the azimuthal and polar frequencies remain entangled, severely limiting the precision of spin measurements. Fisher information analysis demonstrated that the isolation of the nodal frequency shifts the spin estimation into a frequency-limited regime. Over a single coherence time at $200$~Mpc, the statistical error on the dimensionless Kerr spin is constrained to $\sigma_{a} \approx 10^{-2}$. This represents an improvement of up to two orders of magnitude over the equatorial baseline, achieved strictly through the geometric information encoded in the satellite combs.

\par
This multi-comb architecture provides a rigorous null test for the spacetime geometry. Because any planar orbit in a static, spherically symmetric spacetime generates a maximum of three frequency combs, the five-comb structure of an inclined Kerr orbit cannot be replicated by spherically symmetric templates. An analytical lower bound on the waveform mismatch confirmed that even modest orbital inclinations are sufficient to completely rule out static black hole models. Hypothesis testing via simulated data injections verified that the Kerr signal separates decisively from the spherical baseline. By extracting the nodal drift directly from the data stream, inclined zoom-whirl orbits offer a model-independent channel to verify the rotational nature of astrophysical black holes.

%% Also note that the akcnowlodgment environment does not support long amounts of text. If you have a lot of people and institutions to acknowledge, do not use this command. Instead, create a new \section{Acknowledgments}.

\acknowledgments

This work is supported by the National Natural Science Foundation of China (Grant No. 12505060), Fapesq-PB of Brazil, the Fund Project of Chongqing Normal University (Grant Number: 24XLB033), Chongqing Natural Science Foundation General Program (Grant No. CSTB2025NSCQ-GPX1019), and Science and Technology Research Project of Chongqing Municipal Education Commission (Grant Number: KJQN202500563).


%\section*{Data/software/code availability}
%The figures, numerical data, and code supporting this work are available from the authors upon reasonable request.

\bibliographystyle{JHEP}
\bibliography{refs}
\end{document}